\documentclass[prb,aps,twocolumn,floats,amsmath,showpacs]{revtex4}

\usepackage[pdftex]{color,graphicx}
\usepackage{epstopdf}
\usepackage{subfigure}
\usepackage{amssymb}
\usepackage{algorithm}
\usepackage[noend]{algpseudocode}
\usepackage{color,appendix}
\usepackage{tikz,pgfplots,pgfplotstable}
\usetikzlibrary{positioning}

\newcommand{\funD}[2]{{\delta #1\over \delta #2}}

\newcommand\subsetsim{\mathrel{%
  \ooalign{\raise0.2ex\hbox{$\subset$}\cr\hidewidth\raise-0.8ex\hbox{\scalebox{0.9}{$\sim$}}\hidewidth\cr}}}

\begin{document}
\title{RESCU: a Real Space Electronic Structure Method}
\author{Vincent Michaud-Rioux}
\email{Email:vincentm@physics.mcgill.ca}
\author{Lei Zhang}
\email{Email:zhanglei@physics.mcgill.ca}
\author{Hong Guo}
\affiliation{Center for the Physics of Materials, Department of Physics, McGill University, Montreal, Canada H3A 2T8 }

\begin{abstract}
In this work we present RESCU, a powerful MATLAB-based Kohn-Sham density functional theory (KS-DFT) solver. We demonstrate that RESCU can compute the electronic structure properties of systems comprising many thousands of atoms using modest computer resources, e.g. 16 to 256 cores. Its computational efficiency is achieved from exploiting four routes. First, we use numerical atomic orbital (NAO) techniques to efficiently generate a good quality initial subspace which is crucially required by Chebyshev filtering methods. Second, we exploit the fact that only a subspace spanning the occupied Kohn-Sham states is required, and solving accurately the KS equation using eigensolvers can generally be avoided. Third, by judiciously analyzing and optimizing various parts of the procedure in RESCU, we delay the $O(N^3)$ scaling to large $N$, and our tests show that RESCU scales consistently as $O(N^{2.3})$ from a few hundred atoms to more than 5,000 atoms when using a real space grid discretization. The scaling is better or comparable in a NAO basis up to the 14,000 atoms level. Fourth, we exploit various numerical algorithms and, in particular, we introduce a partial Rayleigh-Ritz algorithm to achieve efficiency gains for systems comprising more than 10,000 electrons. We demonstrate the power of RESCU in solving KS-DFT problems using many examples running on 16, 64 and/or 256 cores: a 5,832 Si atoms supercell; a 8,788 Al atoms supercell; a 5,324 Cu atoms supercell and a small DNA molecule submerged in 1,713 water molecules for a total 5,399 atoms. The KS-DFT is entirely converged in a few hours in all cases. Our results suggest that the RESCU method has reached a milestone of solving thousands of atoms by KS-DFT on a modest computer cluster.
\end{abstract}

\pacs{
31.15.E-,    
71.15.-m,    
02.70.Bf,    
31.15.xr,    
}

\maketitle

\section{Introduction}\label{sec:intro}

Density functional theory (DFT)\cite{HK64} based numerical programs are nowadays the standard tool for predicting and understanding the structural and electronic properties of materials that involve many electrons. The idea of treating complicated many-body interactions in real materials by a self-consistent mean field theory appeared in the early days of quantum mechanics. In 1927, Thomas and Fermi proposed a semiclassical model\cite{Thomas27,Fermi27} in which electrons in an external potential are described using only the electronic density. Subsequent calculations are simplified since the complicated many-body wavefunction is avoided. The Thomas-Fermi model was later improved by Dirac who included an exchange energy functional\cite{Dirac30} and by von Weizs\"acker who added a gradient correction to the kinetic energy functional\cite{Weizsacker53}. Nearly four decades later, Hohenberg and Kohn (HK) put DFT on firm theoretical footing by proving that the ground-state expectation values are functionals of the density and that the ground-state density can be calculated by minimizing an energy functional\cite{HK64}. Certain assumptions of the original HK theorems, such as the ground-state non-degeneracy, were later relaxed or eliminated\cite{Levy79}. These theories proved that the ground-state properties of any electronic system can in principle be calculated - if not necessarily understood - without using many-body wave functions. For practical applications, Kohn and Sham (KS) demonstrated that the problem of minimizing the total energy of a system with respect to the electronic density could take the form of a non-interacting electron problem\cite{KS65}. In the KS formulation, the kinetic energy is evaluated via single particle wave functions which is more accurate than using kinetic energy functionals that depend explicitly on the density. KS-DFT allows one to analyze a variety of physical systems and performing a DFT calculation today is all but synonymous to solving the KS equation\cite{KS65}.

Various approaches for solving the KS equation have emerged such as the full potential all-electron methods\cite{Blaha1990,Schwarz2003} and the \emph{ab initio} pseudopotential methods\cite{Pickett1989,Fuchs1999,KresseMix,Vasp2,SIESTA}. In KS-DFT solvers, several bases have been used to express quantum mechanical operators including real space Cartesian grids, finite elements, planewaves, wavelets, numerical atomic orbitals (NAO), Gaussian orbitals, muffin-tin orbitals and some others. The goal is to predict structural and electronic properties of real materials reaching the required accuracy for the given research topic and KS-DFT is playing a prominent role in materials physics and engineering.

At present, a major issue of practical DFT methods is their limited capability of solving material problems involving large number of atoms using a small computer cluster (e.g. 16 to 256 cores). For instance, algorithms implemented in state-of-the-art electronic packages such as VASP\cite{KresseMix,Vasp2} and AbInit\cite{Gonze2009} can comfortably solve systems comprising of a few hundred atoms - but not many thousands on such small computer clusters. With the widening accessibility of supercomputers and the developments of advanced parallel computing algorithms, heroic KS-DFT calculations at the level of 10,000 atoms became possible in recent years, but at the expense of using thousands or even tens of thousands of computing cores\cite{Bottin2008,SIESTA-PEXSI,Levitt14}. Nevertheless, for practical material research and innovation, many small research groups in the world do not have access, cannot afford or simply wish not to use supercomputers. An urgent and very important task is to develop a KS-DFT method that can solve the KS equation without degrading the solution accuracy, at the level of several thousand atoms or more on a small computer cluster. It is the purpose of this work to report and describe such a KS-DFT solver and its associated software implementation.

To see why it is still possible to gain computational efficiency in traditional eigenvalue-based KS-DFT approaches, we note - as others had noted before us\cite{Zhou06-1,Zhou06-2} - that the solution process of the KS-DFT is a self-consistent procedure where one numerically converges the Hamiltonian step by step by solving the KS equation repeatedly and accurately. However, it appears unclear why one has to solve accurately the KS equation for the not-yet-converged Hamiltonian in the intermediary steps. Another observation is that, in the eigensolver-based KS-DFT methods, different parts in the computation scale differently as a function of electron number $N$, some $O(N)$, others $O(N^2)$ and eventually these are dominated by the $O(N^3)$ parts. If one is able to \emph{``delay"} the crossover to $O(N^3)$ scaling, larger systems can potentially be solved using small computers. It turns out that these computational gains can be realized as we present below.

Our KS-DFT method combines NAO and the real space finite-differences plus Chebyshev filtering (CF) technique introduced by Zhou \textit{et al.}\cite{Zhou06-1,Zhou06-2}. We found it is key to generate efficiently a proper initial subspace in the Chebyshev filtering framework, and this is achieved by the use a NAO basis. We advance efficient parallelization, a partial Rayleigh-Ritz (pRR) method for the computation of the density matrix and careful optimization of the solution process, and we have reached our goal of solving solid state physics problems consisting of thousands of atoms using 16 to 256 cores. Our code is called RESCU - which stands for Real space Electronic Structure CalcUlator - and it is implemented in the technical computing platform MATLAB. We use our own MPI and ScaLAPACK interfaces to harness efficiently the computational power of the cores. As such, RESCU combines the vocations of a prototyping code and a production code. In particular, the pRR allows us to compute the single particle density matrix in problems involving an exceedingly large number of electrons by taking advantage of the quasi-minimal property of basis sets built from CF.  In the present paper, we will present the algorithmic and implementation advancements achieved during the development of RESCU. As practical examples, we demonstrate the following KS-DFT calculations: we simulate 5,832 Si atoms (23,328 electrons) on a real space grid, converging the entire KS-DFT calculation using 256 cores for about 5.5 hours; we simulate 4,000 Al atoms (12,000 electrons) on a real space grid, converging the entire KS-DFT calculation using 64 cores for about 5.1 hours; we simulate a supercell consisting of 13,824 Si atoms (55,296 electrons) using a NAO basis, converging the entire calculation using 64 cores for about 6.4 hours; we simulate a supercell consisting of 5,324 Cu atoms (58,564 electrons) using a NAO basis, converging the entire calculation using 256 cores for about 12 hours. We also consider a disordered system consisting of a small DNA molecule submerged in 1,713 water molecules, for a total of 5,399 atoms (14,596 electrons), and converge the entire KS-DFT run in 9.6 hours on 256 cores. These results are compiled in table \ref{tb:largebench} which is found in section \ref{sec:num}. The scaling of the RESCU method is presented going from 16 cores to 256 cores for various tests. Finally, since RESCU is primarily a real space implementation of KS-DFT, it does not require periodicity when dealing with condensed phase materials and can thus easily treat problems involving interfaces, surfaces, defects, disordered materials, etc.

The paper is organized as follow. In section \ref{sec:theory}, we briefly state the fundamentals of DFT and introduce the single particle density matrix theoretical framework which is used throughout this article. In section \ref{sec:algorithms}, we review the state-of-the-art numerical methods for solving the KS equations and recount their advantages and disadvantages. In section \ref{sec:chebfilt}, we describe in some detail the Chebyshev filtering method. In section \ref{sec:PRR}, we present a computational complexity analysis of the Chebyshev filtering method and introduce the partial Rayleigh-Ritz algorithm. We explain how it takes advantage of the Chebyshev filtered basis sets to improve on the standard Rayleigh-Ritz algorithm. In section \ref{sec:imp}, we describe the implementation of the Kohn-Sham DFT solver RESCU. In section \ref{sec:num}, we present different benchmarks of the RESCU code. We provide evidence when the partial Rayleigh-Ritz algorithm achieves significant gains over the standard Rayleigh-Ritz algorithm. We show how to generate a good quality initial subspace efficiently. Finally, we report simulations including thousands of atoms with modest computer resources. Bottlenecks and future direction will be discussion in section \ref{sec:discussion} and \ref{sec:conclusion}.

\section{A brief discussion of DFT}\label{sec:theory}

Before delving into the details of the RESCU method, we briefly discuss KS-DFT in general terms. As mentioned in the introduction, the founding result of DFT is that the Hamiltonian of a system is uniquely determined by its ground-state electronic density\cite{HK64,Levy79}. It follows that the ground-state wave function and the associated expectation values are also determined by the ground-state density, and there exists a universal energy functional of the density which is minimized by the ground-state density\cite{HK64,Levy79}.

In the KS-DFT, the problem of minimizing the energy with respect to the density is mapped to a non-interacting electron problem\cite{KS65}. The KS-DFT formulation made it possible to develop reasonably accurate energy functionals and it became the most successful and widely applied flavor of DFT. In particular, accurate kinetic energy functionals use the Kohn-Sham orbitals and not the density \textit{per se}. The Kohn-Sham equation is usually written as the the following set of equations
\begin{align}
\label{eq:KS1}
\lambda_i\psi_i &= \left(-{1\over 2}\nabla^2 + V_{ext} + V_H + V_{xc}\right)\psi_i\\
\label{eq:rho}
\rho(\mathbf{r}) &= \sum\limits_{i=1}^{\infty} n_{FD}(\lambda_i,\mu)\psi_i^*(\mathbf{r})\psi_i(\mathbf{r})\\
\label{eq:vh}
\nabla^2 V_{H} &= 4\pi\rho\\
\label{eq:vxc}
V_{xc} &= \funD{E_{xc}[\rho]}{\rho}
\end{align}
The density $\rho$ is the sum of the squared norm of the Kohn-Sham wave functions weighted by the Fermi-Dirac distribution. At zero temperature, the only populated states are the $N$ lowest lying states where $N$ is the number of electrons in the system. The Hartree potential $V_H$ may be obtained by solving the Poisson equation, the exchange-correlation potential $V_{xc}$ is defined as the functional derivative of the exchange-correlation energy functional $E_{xc}$ with respect to the density, $n_{FD}$ denotes the Fermi-Dirac distribution and the chemical potential $\mu$ is set such that the number of electrons is $N$. The equations are written in atomic units, we denote the Hamiltonian $\mathbf{H}=-{1\over 2}\nabla^2 + V_{ext} + V_H + V_{xc}$; its dimension, which corresponds to the number of real space grid points or the number of k-space grid points, is $M$ and the eigenvalues are indexed from smallest to largest as follows $\lambda_1<\lambda_2<...\lambda_{M-1}<\lambda_M$. Unlike the Schr\"odinger equation, the KS equation is non-linear since the potential depends on the density which in turn depends on the KS eigenstates. Consequently, the KS equation must be solved by cycling through Eqs.(\ref{eq:KS1}) to (\ref{eq:vxc}) until a fixed point $\rho^*$ is found although other convergence criteria may be used.

We introduce a more flexible framework in which Eqs.(\ref{eq:KS1}) to (\ref{eq:vxc}) are expressed in terms of the single particle density matrix defined as follows
\begin{align}
\label{eq:dmpsi}
\rho(\mathbf{r},\mathbf{r}') &= \sum\limits_{i=1}^{\infty} n_{FD}(\lambda_i,\mu)\psi_i^*(\mathbf{r})\psi_i(\mathbf{r}')\ .
\end{align}
Note that the density is simply the diagonal of the single particle density matrix. The Fermi-Dirac distribution $n_{FD}(\lambda,\mu)$ decays exponentially fast for $\lambda > \mu$. It is thus reasonable to set the occupation number to zero if $n_{FD}(\lambda_i,\mu) < \epsilon$ where $\epsilon$ is some tolerance. Consequently, the number of required Kohn-Sham states $L$ is equal to or slightly larger than the number of electrons $N$ and it is much smaller than the linear dimension of the Hamiltonian matrix $M$. In other words, the single particle density matrix is a low rank matrix. It is convenient to rewrite Eq.(\ref{eq:dmpsi}) using matrix notation. To that end, we define the populated Kohn-Sham eigenspace as
\begin{align}
\label{eq:defPSI}
\boldsymbol\Psi = \left[\psi_1 \psi_2 ... \psi_{L-1} \psi_{L}\right]
\end{align}
where $\psi_i$ satisfies Eq.(\ref{eq:KS1}). The density matrix is then expressed as follows
\begin{align}
\label{eq:DMPSI}
\mathbf{P} = \boldsymbol\Psi n_{FD}(\boldsymbol{\Lambda},\mu)\boldsymbol\Psi^\dagger
\end{align}
where $\boldsymbol{\Lambda}_{ij} = \lambda_i\delta_{ij}$. The Kohn-Sham eigenstates are expressed in terms of basis functions $\{\phi_i\}$ as done in the following equation:
\begin{align}
\label{eq:ksphi}
\psi_j &= \sum\limits_i c_{ij}\phi_i\ ,
\end{align}
which translates as follow in matrix notation:
\begin{align}
\label{eq:ksPHI}
\mathbf{\Psi} &= \mathbf{\Phi}\mathbf{C}\ .
\end{align}
Inserting (\ref{eq:ksPHI}) in (\ref{eq:DMPSI}) we obtain
\begin{align}
\label{eq:RhoPhi}
\mathbf{P} = \mathbf{\Phi}\mathbf{C} n_{FD}(\boldsymbol{\Lambda},\mu)\mathbf{C}^\dagger\mathbf{\Phi}^\dagger\ .
\end{align}
The matrix $\mathbf{C}$ satisfies a generalized eigenvalue equation,
\begin{align}
\label{eq:projGV}
\overline{\mathbf{H}}\mathbf{C} &= \overline{\mathbf{S}}\mathbf{C}\boldsymbol{\Lambda}
\end{align}
where $\overline{\mathbf{H}} = \mathbf{\Phi}^\dagger \mathbf{H} \mathbf{\Phi}$, and $\overline{\mathbf{S}} = \mathbf{\Phi}^\dagger \mathbf{\Phi}$ is the overlap matrix. Since the overlap matrix is symmetric positive definite, it is possible to calculate its Cholesky decomposition $\overline{\mathbf{S}} = \overline{\mathbf{U}}^T\overline{\mathbf{U}}$. Eq.(\ref{eq:projGV}) is usually solved by reducing the generalized eigenvalue problem to a standard eigenvalue problem
\begin{align}
\label{eq:projEV}
\hat{\overline{\mathbf{H}}}\hat{\mathbf{C}} &= \hat{\mathbf{C}}\boldsymbol{\Lambda}
\end{align}
where $\hat{\overline{\mathbf{H}}} = \overline{\mathbf{U}}^{-T}\overline{\mathbf{H}}\overline{\mathbf{U}}^{-1}$ and $\hat{\mathbf{C}} = \overline{\mathbf{U}}\mathbf{C}$.
Plugging Eq.(\ref{eq:projEV}) in (\ref{eq:RhoPhi}) we obtain:
\begin{align}
\label{eq:RhoPhi2}
\mathbf{P} &= \mathbf{\Phi}\overline{\mathbf{U}}^{-1} n_{FD}(\hat{\overline{\mathbf{H}}},\mu)\overline{\mathbf{U}}^{-T}
\mathbf{\Phi}^\dagger
\end{align}
It appears that as long as $\boldsymbol\Psi$ is a subspace of $\mathbf{\Phi}$, the density matrix is unchanged and so is the electronic density. Note that the chemical potential $\mu$ satisfies
\begin{align}
\label{eq:DM2N}
N = Tr\left[n_{FD}(\hat{\overline{\mathbf{H}}},\mu)\right]\ .
\end{align}
We shall refer to the quantity $\overline{\mathbf{P}}$, defined in Eq. (\ref{eq:projDM}) below, as the projected density matrix even though $\overline{\mathbf{P}} \neq \mathbf{\Phi}^\dagger \mathbf{P} \mathbf{\Phi}$.
\begin{align}
\label{eq:projDM}
\overline{\mathbf{P}} &= \overline{\mathbf{U}}^{-1} n_{FD}(\hat{\overline{\mathbf{H}}},\mu)\overline{\mathbf{U}}^{-T}\ .
\end{align}

\section{Practical algorithms for solving the KS equation}\label{sec:algorithms}

The purpose of this section is to review briefly practical algorithms used in state-of-the-art and recent DFT solvers, with the focus on elucidating where and how one may achieve speed-ups so that larger systems can be solved by KS-DFT using a modest computer.

Using the definitions introduced in section \ref{sec:theory}, we now reformulate the KS equation as described in the generic Kohn-Sham solver Algorithm \ref{alg:KSgeneric} below. Firstly, the electronic density is initialized and the dual Hamiltonian generated. The density is often initialized using the isolated atom densities but other choices, a uniform density for example, are viable in certain systems. A subspace $\mathbf{\Phi}^k$ which spans approximately the occupied Kohn-Sham subspace $\mathbf{\Psi}$ is generated. Then the Hamiltonian and identity operators are projected onto the subspace $\mathbf{\Phi}^k$. The projected density matrix is then evaluated, typically by solving the matrix equation (\ref{eq:projGV}) or (\ref{eq:projEV}). Next, the density is obtained from the diagonal of the real space single particle density matrix and the Hamiltonian is updated by solving the Poisson equation and evaluating the exchange-correlation potential. This step is generally preceded by a mixing of the density or followed by a mixing of the potential. Finally, the convergence of the density and possibly other quantities are monitored.

\begin{algorithm}
\caption{Generic Kohn-Sham solver}
\label{alg:KSgeneric}
\begin{algorithmic}
\Procedure{GenericSolver}{$\delta$}
\State Initialize $\rho_0$, $\mathbf H[\rho_0]$
\While{$\epsilon > \delta$ or $k < k_{max}$}
\State Compute a subspace $\boldsymbol\Phi^k$ which spans $\boldsymbol\Psi^k$
\State Compute the projected Hamiltonian $\overline{\mathbf{H}}^k$ and the overlap matrix $\overline{\mathbf{S}}^k$
\State Compute the projected density matrix $\overline{\mathbf{P}}^k$
\State Compute the density $\rho^{k}(\mathbf{r})=\mathbf{P}^k(\mathbf{r},\mathbf{r})$
\State Compute $\mathbf H = \mathbf H[\rho_{k}]$
\State Calculate $\epsilon = \lVert\rho_{k} - \rho_{k-1}\rVert$, $\epsilon = \lVert\mathbf H[\rho_{k}] - \mathbf H[\rho_{k-1}]\rVert$
\EndWhile
\State \textbf{return} $\rho_{k}$
\EndProcedure
\end{algorithmic}
\end{algorithm}

Many currently used DFT codes fit in the framework established in Algorithm \ref{alg:KSgeneric}. They generally differ in how to calculate the subspace $\mathbf{\Phi}^k$ and how to compute the projected density matrix $\overline{\mathbf{P}}^k$: these are the foci of the most recent algorithmic advancements and probably those to come as we shall mention later. We now discuss how particular DFT methods translate in the above picture.

The procedure executed by state-of-the-art DFT solvers is summarized in Algorithm \ref{alg:KSDiag} below. The eigenvectors of the Kohn-Sham Hamiltonian are calculated directly which makes the rest of the procedure simple. The Kohn-Sham states diagonalize the Hamiltonian such that $\overline{\mathbf{H}}^k$ is diagonal and the overlap matrix is the identity $\mathbf{I}$ by virtue of the orthogonality of the eigenvectors. Calculating the Fermi-Dirac operator $n_{FD}(\hat{\overline{\mathbf{H}}}^k,\mu) = n_{FD}(\boldsymbol{\Lambda}^k,\mu)$ is then trivial. At zero temperature, $N$ Kohn-Sham states are required if there are $N$ electrons in the system ($N/2$ if there is spin degeneracy). When using thermal smearing, more states are required since the states whose energy is close to $\mu$ are fractionally occupied. As mentioned above, the density matrix has a low rank since $n_{FD}(\lambda,\mu)$ decays exponentially fast for $\lambda>\mu$. It is thus generally sufficient to compute $L = N + N_{buf}$ KS states where $N_{buf}$ is a modest number. These $L$ KS states can be thought of as forming a KS basis that diagonalize the KS Hamiltonian, and this Kohn-Sham basis is deemed quasi-minimal since $L\simeq N$. The number of Kohn-Sham basis functions is typically smaller than the dimensionality of the discretized Hamiltonian by \emph{a few orders of magnitude} and therefore it is advantageous to use partial diagonalization methods such as the Arnoldi algorithm or the Lanczos algorithm to compute the required eigenvectors. These algorithms are implemented in established software libraries such as ARPACK\cite{Lehoucq98} and TRLAN\cite{Wu99} which are used by many DFT solvers.

\begin{algorithm}
\caption{Diagonalization Kohn-Sham solver}\label{alg:KSDiag}
\begin{algorithmic}
\Procedure{DiagSolver}{$\delta$}
\State Initialize $\rho_0$, $\mathbf H[\rho_0]$
\While{$\epsilon > \delta$ or $k < k_{max}$}
\State Compute the occupied Kohn-Sham subspace $\boldsymbol\Psi^k$
\State The projected Hamiltonian is $\boldsymbol\Lambda^k$ and the overlap matrix $\mathbf{I}$
\State The projected density matrix is $n_{FD}(\boldsymbol{\Lambda}^k,\mu)$
\State Compute the density $\rho^{k+1}(\mathbf{r})=\mathbf{P}^k(\mathbf{r},\mathbf{r})$
\State Compute $\mathbf H = \mathbf H[\rho_{k}]$
\State Calculate $\epsilon = \lVert\rho_{k} - \rho_{k-1}\rVert$, $\epsilon = \lVert\mathbf H[\rho_{k}] - \mathbf H[\rho_{k-1}]\rVert$
\EndWhile
\State \textbf{return} $\rho_{k+1}$
\EndProcedure
\end{algorithmic}
\end{algorithm}

Even sparse diagonalization techniques are very computationally demanding if a lot of occupied states must be computed. As already mentioned at the end of the last section, the Kohn-Sham states are actually not necessary and a subspace which approximately spans the occupied Kohn-Sham subspace may solve the KS equation. For example, basis sets such as atomic orbitals\cite{FHI,Conquest,GPAW,SIESTA} and Gaussian orbitals\cite{g09} have been used extensively in the DFT community. The procedure using a predefined basis set is summarized in Algorithm \ref{alg:KSNAO}.

The main disadvantage of such methods is the difficulty to systematically augment the basis to improve the simulation accuracy or validate convergence. Many research groups have put forward methods to generate basis sets that can achieve a systematic convergence for most elements from H to Rn\cite{FHI, SIESTA, Junquera2001, Ozaki2004, Frisch2003, Ahlrichs1992}. Well established DFT codes using atom-centered basis functions such as SIESTA\cite{SIESTA}, FHI-AIMS\cite{FHI} or Gaussian\cite{g09} provide tested basis sets and have been used extensively by researchers to study physical systems with a variety of chemical environments. Nevertheless, some systems
(e.g. certain metals, dense structures, solids with large coordination numbers)
may require special treatment where new basis functions must be generated and tested. This is in contrast with the procedure described in Algorithm \ref{alg:KSDiag} where the Kohn-Sham states accuracy is only determined by the underlying numerical grid, which makes the convergence with respect to the basis set relatively more transparent and straightforward.

On the other hand, predefined basis sets have many computational advantages. In the scheme of Algorithm \ref{alg:KSNAO}, the subspace needs not be updated at every step and hence it is generated ahead of the self-consistent loop. Other quantities such as the projected kinetic energy matrix, the projected non-local ionic potential matrix and the overlap matrix are also computed ahead of the iterative process. Only the diagonal part of the Hamiltonian corresponding to the Hartree and exchange-correlation potentials has to be updated and projected onto the subspace $\mathbf{\Phi}$. Among other advantages, predefined basis sets are often localized by design and the projection of diagonal operators scales linearly with respect to system size and has a relatively cheap computational cost. The basis functions may also have spherical symmetry which makes it possible to perform certain integrals analytically. The main bottleneck is the computation of the projected density matrix which is usually obtained by diagonalizing the reduced projected Hamiltonian $\hat{\overline{\mathbf{H}}}^k$. Whereas these basis sets are not minimal, their dimension is generally a modest multiple of the number of electrons. For systems with less than a thousand atoms or so, these methods are competitive as the matrix $\overline{\mathbf{H}} - \lambda \overline{\mathbf{S}}$ is directly invertible or diagonalizable. In larger systems, it becomes crucial that the projected Hamiltonian matrix be made as small as possible, and these methods become less competitive.

\begin{algorithm}
\caption{Orbital Kohn-Sham Solver}\label{alg:KSNAO}
\begin{algorithmic}
\Procedure{OrbitalSolver}{$\delta$}
\State Initialize $\rho_0$, $\mathbf H[\rho_0]$
\While{$\epsilon > \delta$ or $k < k_{max}$}
\State The subspace $\boldsymbol\Phi$ is constant
\State Compute the projected Hamiltonian $\overline{\mathbf{H}}^k$; the overlap matrix $\overline{\mathbf{S}}$ is constant
\State Compute the projected density matrix $\overline{\mathbf{P}}^k$
\State Compute the density $\rho^{k+1}(\mathbf{r})=\mathbf{P}^k(\mathbf{r},\mathbf{r})$
\State Compute $\mathbf H = \mathbf H[\rho_{k}]$
\State Calculate $\epsilon = \lVert\rho_{k} - \rho_{k-1}\rVert$, $\epsilon = \lVert\mathbf H[\rho_{k}] - \mathbf H[\rho_{k-1}]\rVert$
\EndWhile
\State \textbf{return} $\rho_{k+1}$
\EndProcedure
\end{algorithmic}
\end{algorithm}

A way around this issue is to use polynomial approximations of the Fermi-Dirac operator or other operators simulating its effect. Goedecker and Colombo have used polynomial approximations of the Fermi operator\cite{Goedecker94}; Goedecker and Teter have used Chebyshev polynomial approximations of the complementary error function\cite{Goedecker95} to simulate the action of the Fermi operator; Jay \textit{et al}. used Chebyshev-Jackson polynomial expansions of the Heaviside function\cite{Jay99}; etc. These techniques are free of diagonalization but have bottlenecks and limitations of their own. Polynomial approximations work only at finite temperature since the Fermi-Dirac distribution is discontinuous at $T=0$. Another disadvantage is that the degree of the polynomial must scale as $\mathcal{O}({\beta\sigma})$, where $\beta$ is the inverse temperature and $\sigma$ is the valence spectral width, to achieve a given accuracy. The occupied part of the energy spectrum usually spans many eV and even tens of eV whereas room temperature corresponds to an energy of 0.025 eV - these very different energy scales demand very high order expansions.

It is also possible to compute the density matrix from single particle Green's function using the following formula:
\begin{align}
\label{eq:G2DM}
\mathbf{P} = {1\over 2\pi i}\int\limits_{-\infty}^{\mu}d\lambda \mathbf{G}(\lambda)
\end{align}
where $\mathbf{G} = \left( \lambda - \mathbf{H}\right)^{-1}$, as proposed by Baroni and P. Giannozzi\cite{Baroni92}. Like polynomial expansions, rational approximations generally require a lot of terms to achieve a decent accuracy. The issue has been addressed by Lin and coworkers who developed a multipole expansion method which scales as $\mathcal{O}\log({\beta\sigma})\log(\log({\beta\sigma}))$\cite{Lin09-1,Lin09-2}. Their method was combined with the parallel selective inversion algorithm developed by Lin \textit{et al}.\cite{Lin11-1,Lin11-2} to perform electronic structure calculations of systems comprising thousands of atoms\cite{Lin13,Lin14}. The rational expansion of the Fermi-Dirac operator leads to an inverse intensive method which has some disadvantages compared with diagonalization techniques. It is usually more difficult to achieve a good load balance and memory distribution in inverse algorithms. Moreover, the complexity worsens as the dimensionality of the system increases from one dimension (1D) to three dimensions (3D). This is due to the filling of the matrix factors which becomes problematic in 3D problems from both memory and processing perspectives. If the inverse is available, spectrum slicing and shift-and-invert eigensolvers may also be used to find a large number of eigenvectors efficiently\cite{Aktulga14}.


Having briefly reviewed the various algorithms in practical KS-DFT implementations, we present a method which focuses on building a quasi-minimal basis set in the following subsection. The adaptive basis set is constructed via the Chebyshev filtering technique first applied by Zhou, Chelikowsky, Saad and coworkers\cite{Zhou06-1,Zhou06-2,Saad06,Cheli12}. The workflow is presented in Algorithm \ref{alg:KSCFSI}. First, the basis set from the previous iteration is refined using Chebyshev filtering. A Chebyshev filter is an operator which is expressed as a first kind Chebyshev polynomial of the Hamiltonian matrix. It is easily evaluated as it only requires matrix-vector products.
The size of the subspace is usually comparable to $N$ and is thus significantly smaller than the size of fixed basis sets (e.g. Gaussian orbitals or atomic orbitals). It can also be systematically refined and its accuracy is only limited by the underlying grid resolution. The downside is naturally that its computation comes at a cost and the resulting subspace is generally dense such that the memory requirement scales as $\mathcal{O}(N^2)$. The scheme is generally more costly than DFT calculations using orbital bases but much lighter than the plain diagonalization schemes. RESCU implements both Algorithms \ref{alg:KSNAO} for atomic orbitals and Algorithm \ref{alg:KSCFSI} for real space grids.

\begin{algorithm}
\caption{CFSI Kohn-Sham Solver}\label{alg:KSCFSI}
\begin{algorithmic}
\Procedure{CFSISolver}{$\delta$}
\State Initialize $\rho_0$, $\mathbf H[\rho_0]$
\While{$\epsilon > \delta$ or $k < k_{max}$}
\State Compute a subspace $\boldsymbol\Phi^k = T_n(\mathbf{H})\boldsymbol\Phi^{k-1}$ using the Chebyshev filtering
\State Compute the projected Hamiltonian $\overline{\mathbf{H}}^k$ and the overlap matrix $\overline{\mathbf{S}}^k$
\State Compute the projected density matrix $\overline{\mathbf{P}}^k$
\State Compute the density $\rho^{k+1}(\mathbf{r})=\mathbf{P}^k(\mathbf{r},\mathbf{r})$
\State Compute $\mathbf H = \mathbf H[\rho_{k}]$
\State Calculate $\epsilon = \lVert\rho_{k} - \rho_{k-1}\rVert$, $\epsilon = \lVert\mathbf H[\rho_{k}] - \mathbf H[\rho_{k-1}]\rVert$
\EndWhile
\State \textbf{return} $\rho_{k+1}$
\EndProcedure
\end{algorithmic}
\end{algorithm}



\section{Chebyshev filtering}\label{sec:chebfilt}

\begin{figure}
\centering
\includegraphics[width=\linewidth]{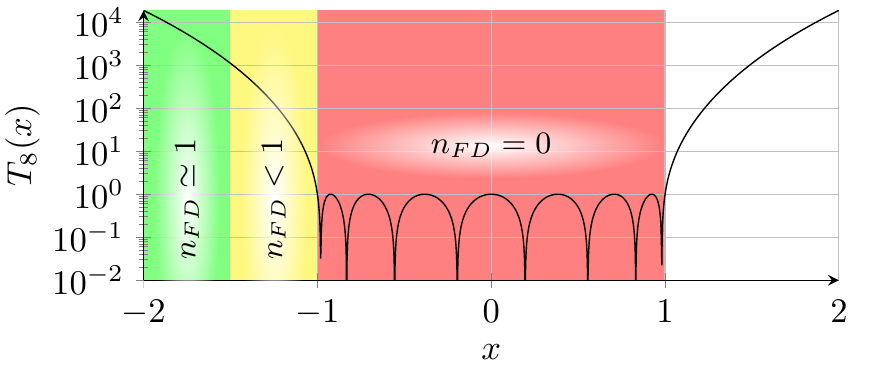}
\caption{8$^{th}$ degree Chebyshev polynomial of the first kind $T_8(x)$ as a function of x. The y axis is logarithmic. The green region is mapped to fully occupied Kohn-Sham states. The yellow region is mapped to the fractionally occupied states (mostly unoccupied states). The [-1,1] interval is mapped to the unoccupied Kohn-Sham states. $|T_8(x)|\leq1$ in the interval [-1,1] and hence applications of the Chebyshev filter suppress the unoccupied components. The interval $[1,\infty]$ is mapped above the spectrum and hence do not contribute.}
\label{fig:T8Plot}
\end{figure}

In this section, we describe the Chebyshev filtering procedure introduced by Zhou \textit{et al}. for KS-DFT calculations\cite{Zhou06-1} and its application in the RESCU method. Already in 2006, the simulation of a Si$_{9041}$H$_{1860}$ nanocluster was reported in Ref.\onlinecite{Zhou06-2} using this method plus computational acceleration by symmetry considerations. The technique concentrates on building a subspace $\boldsymbol\Phi$ which spans $\boldsymbol\Psi$ as defined in Eq.(\ref{eq:defPSI}). A suitable approximation for $\boldsymbol\Psi$ is generated and thereafter rotated toward $\boldsymbol\Psi$ only using Hamiltonian-subspace products. We denote the subspace at the $k^{th}$ self-consistent iteration $\boldsymbol\Phi^k$. To illustrate how this works, we shall refer to Fig.\ref{fig:T8Plot}. The dimension of $\boldsymbol\Phi^k$ is $L = N_{oc} + N_{fr}$, where $N_{oc}$ is the number of fully occupied states ($n_{FD} > 1-\epsilon$) and $N_{fr}$ is the number of fractionally occupied states ($1 -\epsilon \geq n_{FD} \geq 0$).

Consider a vector $\phi\in\mathbb{R}^M$. Since the Kohn-Sham eigenvector basis is complete we can write
\begin{align}
\phi = \sum a_i\psi_i\ .
\end{align}
If we apply a spectral filter $T_n(\mathbf{H})$ to $\phi$, we change the composition of the vector $\phi$ in the following way:
\begin{align}
T_n(\mathbf{H})\phi &= \sum a_iT_n(\mathbf{H})\psi_i \\
&= \sum a_iT_n(\lambda_i)\psi_i\ .
\end{align}
Suppose that $T_n(\lambda_i) \gg T_n(\lambda_j)$ for $i \in \{1,...,L\}$ and $j \in \{L+1,...,M\}$, then $T_n(H)\phi$ has a much larger overlap with $\mathbf{\Psi}$ than $\phi$. Applying such a filter to a whole subspace $\boldsymbol\Phi^k$ will result in a steering of $\boldsymbol\Phi^k$ toward $\boldsymbol\Psi$. For reasons evoked previously, we would like to avoid diagonalization and inversion of the Hamiltonian so that polynomial filters are an evident choice. We seek a polynomial that assumes large values in the interval $[\lambda_{1}, \lambda_{L}]$ and small values in the interval $[\lambda_{L+1}, \lambda_{M}]$. Many polynomials satisfy this property but the Chebyshev polynomials of the first kind (denoted $T_n$ where $n$ is the degree) have the minimal $\infty$-norm on the interval $[-1,1]$ among monic polynomials, and they grow exponentially in the degree $n$ outside $[-1,1]$ making them the ideal candidate. The 8$^{th}$ order Chebyshev polynomial of the first kind $T_8$ is plotted on a logarithmic scale in Fig. \ref{fig:T8Plot}. The polynomial $\infty$-norm is bounded by 1 in the interval $[-1,1]$ and strictly greater than 1 outside the interval $[-1,1]$.

In general, the unoccupied energies do not correspond to the interval $[-1,1]$. In order to use $T_n$ as a filter, we must apply an affine transformation which maps the interval $[-\infty,-1]$ to $[-\infty, \lambda_{L}]$ and $[-1,1]$ to $[\lambda_{L+1}, \lambda_{M}]$. In this way, the components of the occupied spectrum are assuredly magnified with respect to the components of the unoccupied spectrum. This requires the knowledge of the lower and upper bounds $\lambda_{L}$ and $\lambda_{M}$. The lower bound of the unoccupied spectrum $\lambda_{L}$ can be estimated from the largest Ritz value of $\boldsymbol\Phi^k$ which is easily obtainable. The upper bound of the spectrum can be estimated from a few step of the Lanczos algorithm. Estimates for the eigenvalue errors are derived in Templates for the solution of algebraic eigenvalue problems\cite{Bai00} and Zhou has studied the accuracy and robustness of a few estimators for the upper bound of the spectrum\cite{Zhou11}.

Now, consider a system of $N$ electrons at zero temperature such that $N_{oc} = N$. The rate of convergence is roughly $T_n(\lambda_N)$ since $|T_n(\lambda_i)| \geq |T_n(\lambda_N)| \geq 1 \geq |T_n(\lambda_j)|$ for $i\in\{1,...,N\}$ and $j\in\{L+1,...,M\}$. The occupied part of the spectrum is thus magnified by at least $T_n(\lambda_N)$ at every filter application with respect to the unoccupied spectrum. However, if $\lambda_N \simeq \lambda_{L+1}$ then $|T_n(\lambda_N)|\gtrsim 1 \gtrsim |T_n(\lambda_{N+1})| $ and the convergence rate $T_n(\lambda_N)\simeq 1$ is disappointing. It is thus crucial to include enough fractionally occupied states in the subspace to separate the occupied part of the spectrum and the unoccupied part of the spectrum. The fractionally occupied Kohn-Sham states converge slowly but it does not matter since they contribute little if at all to the electronic density. The occupied, fractionally occupied and unoccupied sections of the spectrum are represented by the green, yellow and red region respectively in Fig.\ref{fig:T8Plot}. The white region does not map to any state since the largest eigenvalue is mapped below 1. A pseudocode detailing the application of a Chebyshev filter is found in Algorithm \ref{alg:CF1}. There, we take advantage of the fact that Chebyshev polynomials of the first kind obey the recursive relation $T_{n+1}(x) = 2xT_n(x) - T_{n-1}(x)$.  Alternatively, the coefficients can be directly computed using the following formula\cite{MWTn}
\begin{align}
T_n(x) & = \tfrac{n}{2} \sum_{k=0}^{\left \lfloor \frac{n}{2} \right \rfloor}\left(-{1\over 4}\right)^k 2^n\frac{(n-k-1)!}{k!(n-2k)!} x^{n-2k}
\end{align}
which allows the filtering implementation to use one less temporary vector.

\begin{algorithm}
\caption{Chebyshev Filtering}\label{alg:CF1}
\begin{algorithmic}
\Procedure{ChebFilter}{$n$, $\lambda_M$, $\lambda_{L}$, $\psi_0$, $\mathbf H$}
\State Compute the affine transformation parameters $e=(\lambda_M-\lambda_{L})/2$ and $c=(\lambda_M+\lambda_{L})/2$.
\State $\psi_1 = (\mathbf H\psi_0-c\psi_0)/e$
\For {$i=2,...,n$}
\State $\psi_2 = 2(\mathbf H\psi_1-c\psi_1)/e-\psi_0$
\State $\psi_0 = \psi_1$, $\psi_1 = \psi_2$
\EndFor
\State \textbf{return} $\psi_2$
\EndProcedure
\end{algorithmic}
\end{algorithm}

If $N_{fr}$ was equal to 0, we would only need to orthonormalize the subspace $\mathbf{\Phi}^k$ and the projected density matrix $\overline{\mathbf{P}}$ could be assumed to be the identity and therefore the density would then be trivial to evaluate. This is because the density is invariant under unitary transformations of the occupied Kohn-Sham subspace. However, using extra Kohn-Sham states is generally necessary to obtain a robust convergence for  reasons we just mentioned and the density must not include the contribution from the unoccupied states. For very large systems comprising tens of thousands of electrons, certain quantities such as the total energy may not be affected significantly by including a few unoccupied states in the density. Whether this is true depends on the problem and the precision target but in general Chebyshev filtering is followed by the Rayleigh-Ritz procedure in order populate the Kohn-Sham states correctly. The Rayleigh-Ritz procedure is summarized in Algorithm \ref{alg:RR}. It is essentially a procedure that orthonormalizes the subspace $\mathbf{\Psi}^k$ and computes the projected density matrix. It scales as $\mathcal{O}(N^3)$ and it is the main bottleneck in large scale computations in the Chebyshev filtering scheme. Note that it is possible to first orthonormalize $\boldsymbol\Phi^k$ and then solve a standard eigenvalue problem or use the non-orthonormal $\boldsymbol\Phi^k$ and solve a generalized eigenvalue problem. We have observed that the latter is generally slightly faster overall.

\begin{algorithm}
\caption{Rayleigh-Ritz procedure}\label{alg:RR}
\begin{algorithmic}
\Procedure{RayleighRitz}{$\mathbf H$,$\boldsymbol\Phi$}
\State Compute $\overline{\mathbf{H}} = \boldsymbol\Phi^\dagger \mathbf H \boldsymbol\Phi$
\If{$\boldsymbol\Phi^\dagger \boldsymbol\Phi \neq \mathbf{I}$}
\State Compute $\overline{\mathbf{S}} = \boldsymbol\Phi^\dagger \boldsymbol\Phi$
\Else
\State $\overline{\mathbf{S}} = \mathbf{I}$
\EndIf
\State Diagonalize $\overline{\mathbf{H}}\mathbf{C} = \overline{\mathbf{S}}\mathbf{C}\boldsymbol\Lambda$
\State Compute $\boldsymbol\Phi = \boldsymbol\Phi\mathbf{C}$
\State Compute $\overline{\mathbf{P}} = n_{FD}(\boldsymbol\Lambda,\mu)$
\State \textbf{return} $\boldsymbol\Phi,\overline{\mathbf{P}}$
\EndProcedure
\end{algorithmic}
\end{algorithm}

The Chebyshev filtering technique introduced above can be used with orthonormal discretization schemes such as finite-differencing and plane-waves. It may also be used on the projected Hamiltonian even in the case where the overlap matrix is not the identity. Implementations using Chebyshev filtering with non-orthonormal basis sets such as finite-elements\cite{Motamarri13,Motamarri14}, projector-augmented waves (PAW)\cite{Levitt14} and full-potential linearized augmented planewaves (FLAPW)\cite{Berljafa14} have been reported in the literature. We will describe how the technique also benefits basis sets methods such as NAO. As mentioned above, the NAO basis is static and localized. This leads to a sparse representation for $\overline{\mathbf{H}}$ and $\overline{\mathbf{S}}$ but their size is not minimal such that the eigenvalue problem of the Rayleigh-Ritz procedure is rather large. The Chebyshev filtering technique can construct and maintain an eigensubspace $\mathbf C^k$ which satisfies, to a good approximation, Eq.(\ref{eq:projGV}). The matrix pencil $\overline{\mathbf{H}} - \lambda\overline{\mathbf{S}}$ cannot be used directly so that one must transform the generalized problem to a standard one. It is crucial to preserve to a point the sparsity of the matrix pencil since otherwise the required matrix operations do not have a significant advantage over dense diagonalization algorithm which are based on QR-decompositions. One option is to rewrite Eq.(\ref{eq:projGV}) as follows:
\begin{align}
\label{eq:SH..L}
\overline{\mathbf{S}}^{-1}\overline{\mathbf{H}}\mathbf{C} &= \mathbf{C}\boldsymbol{\Lambda}
\end{align}
and to apply a filter $T_n(\overline{\mathbf{S}}^{-1}\overline{\mathbf{H}})$. The operator $\overline{\mathbf{S}}^{-1}\overline{\mathbf{H}}$ is no more Hermitian but in principle Eqs.(\ref{eq:projGV}) and (\ref{eq:SH..L}) are equivalent and this should pose no problem. One issue is that $\overline{\mathbf{S}}^{-1}$ is generally dense and hence the matrix-vector products are computationally costly. In quasi-1D or quasi-2D systems, the Cholesky factor $\overline{\mathbf{U}}$ may still be relatively sparse depending on the system and the ordering of the overlap matrix. In this case, we suggest considering Eq.(\ref{eq:projEV}) instead. The reduced Hamiltonian $\hat{\overline{\mathbf{H}}} = \overline{\mathbf{U}}^{-T}\overline{\mathbf{H}}\overline{\mathbf{U}}^{-1}$ can be used in the filter to compute $\hat{\mathbf{C}}^k$ which yields $\mathbf{C}^k$. Each product then necessitates solving two triangular systems of equations and one symmetric matrix product. In conclusion, the efficiency of the Chebyshev filtering method vitally depends on the capacity to invert the overlap matrix in the context of NAO. There is so far no versatile and effective method to solve this issue.

\section{The Partial Rayleigh-Ritz procedure}\label{sec:PRR}

We begin this section by analyzing the computational complexity of the different operations performed during the self-consistent procedure in KS-DFT.

Applying a Chebyshev filter consists essentially in matrix products and the procedure scales as $\mathcal{O}(MN+N^2)$ where $M$ is the size of the discrete Hamiltonian $\mathbf H$ (i.e. the number of grid points) and $N$ is the number of occupied Kohn-Sham states (recall $L \simeq N$). The $\mathcal{O}(MN)$ scaling comes from applying the Laplacian and the $\mathcal{O}(N^2)$ term comes from applying the Kleinman-Bylander projectors in dealing with the nonlocal pseudopotentials\cite{KB82}.

Next, the Rayleigh-Ritz procedure scales as $\mathcal{O}(MN^2+N^3)$ but its computational cost does not dominate until quite large system sizes as we shall demonstrate in Section \ref{sec:num}. We further decompose the complexity of the Rayleigh-Ritz procedure in the following four operations: subspace orthonormalization, Hamiltonian (and identity) projection, eigenvalue solution and computation of the Ritz vectors. The scaling of the most computationally expensive steps in solving the Kohn-Sham equations is displayed in Table \ref{tb:KSscaling}. The non-orthonormality of $\boldsymbol\Phi$ following the Chebyshev filtering procedure can be taken into account by the Rayleigh-Ritz procedure, and therefore the cost of orthonormalization can be absorbed in the diagonalization cost. The complexity for computing $\overline{\mathbf{H}}$ and $\overline{\mathbf{S}}$, and computing $\boldsymbol\Phi := \boldsymbol\Phi\mathbf{C}$ is $\mathcal O(MN^2)$. The former is more computationally expensive since it is a $(N\times M)\times(M\times N)$ matrix product (where $M \gg N$) whereas the latter is a $(M\times N)\times(N\times N)$ matrix product. We identify the computation of $\overline{\mathbf{H}}$ and $\overline{\mathbf{S}}$, the eigenvalue problem $\overline{\mathbf{H}}\mathbf{C}=\overline{\mathbf{S}}\mathbf{C}\boldsymbol\Lambda$ and the computation of the Ritz vectors as the three principal bottlenecks in a large scale KS-DFT computation.

\begin{table}
\begin{tabular}{ll}
\hline
\hline
Procedure & Scaling \\
\hline
Compute $\boldsymbol\Phi := T_n(\mathbf H)\boldsymbol\Phi$ & $\mathcal O(MN)$ \\

Orthonormalize $\boldsymbol\Phi$ & $\mathcal O(MN^2)$ \\

Compute $\overline{\mathbf{H}}$ and $\overline{\mathbf{S}}$ & $\mathcal O(MN^2)$ \\

Solve $\overline{\mathbf{H}}\mathbf{C}=\overline{\mathbf{S}}\mathbf{C}\boldsymbol\Lambda$ & $\mathcal O(N^3)$ \\

Compute $\boldsymbol\Phi := \boldsymbol\Phi\mathbf{C}$ & $\mathcal O(MN^2)$ \\
\hline
\hline
\end{tabular}
\caption{List of the most computationally expensive procedures in solving the Kohn-Sham equations and the associated computational complexities. $M$ is the size of the Hamiltonian matrix  and $N$ is the number of occupied Kohn-Sham states.}
\label{tb:KSscaling}
\end{table}

The obvious way to address the first and third bottlenecks in Table \ref{tb:KSscaling} is to construct a localized basis for $\boldsymbol\Phi^k$ which leads to a sparse matrix representation. In atomic orbital methods, $\boldsymbol\Phi^k$ is sparse but has known limitations in accuracy due to the inflexible nature of the basis set. In addition, even if $\boldsymbol\Phi_{k-1}$ is localized, $\boldsymbol\Phi_k$ is not sparse in general since the Chebyshev filtering procedure fills in the matrix. In Ref.\onlinecite{Motamarri14}, Motamarri \textit{et al}. use the localization technique introduced by Cervera\cite{Cervera09} to build a localized basis for the Chebyshev filtered subspace and the efficiency relies on the possibility to maintain a sparse basis. Their work shows that finite-elements are most appropriate to exploit the advantages of both Chebyshev filtering and basis localization. This is not possible if high-order finite-differences or plane waves are employed to compute the derivatives. When using high-order finite-differences, the density of the subspace matrix increases rapidly with the degree of the Chebyshev filter due to the far reaching high-order stencils.

Here, we address the second bottleneck in Table \ref{tb:KSscaling} by showing that it is unnecessary to fully diagonalize $\hat{\overline{\mathbf{H}}}$ in order to populate the Kohn-Sham states correctly. Suppose that $N_{oc}$ Kohn-Sham states are fully occupied and that $N_{fr}$ are fractionally occupied. We claim that only the $N_{fr}$ largest eigenpairs of $\hat{\overline{\mathbf{H}}}$ are actually required for the KS-DFT, and this method is named the partial Rayleigh-Ritz (pRR) procedure. To see this, consider
\begin{align}
\label{eq:subspaceFP}
\hat{\mathbf{C}} &= \left[\hat{\mathbf{C}}_{oc} \hat{\mathbf{C}}_{fr}\right]\\
\boldsymbol{\Lambda} &= \left[\begin{array}{cc}
\boldsymbol{\Lambda}_{oc} & \mathbf{0} \\
\mathbf{0} & \boldsymbol{\Lambda}_{fr}
\end{array} \right]
\end{align}
where $\hat{\mathbf{C}}$ is the matrix of eigenvectors of $\hat{\overline{\mathbf{H}}}$.
Then
\begin{align}
n_{FD}(\boldsymbol{\Lambda}) = \left[\begin{array}{cc}
\mathbf{I}_{oc} & \mathbf{0} \\
\mathbf{0} & n_{FD}(\boldsymbol{\Lambda}_{fr})
\end{array} \right]\\
= \mathbf{I} + \left[\begin{array}{cc}
\mathbf{0} & \mathbf{0} \\
\mathbf{0} & n_{FD}(\boldsymbol{\Lambda}_{fr}) - \mathbf{I}_{frac}
\end{array} \right]
\end{align}
where $\mathbf{I}_{oc/fr}$ is a $N_{oc/frac}\times N_{oc/fr}$ identity matrix. Using the last equation and the fact that $\hat{\mathbf{C}}$ is unitary, Eq.(\ref{eq:projDM}) can be transformed into
\begin{align}
\label{eq:projDMPart}
\overline{\mathbf{P}} &= \overline{\mathbf{U}}^{-1}\left(\mathbf{I} + \hat{\mathbf{C}}_{fr}\left[n_{FD}(\boldsymbol{\Lambda}_{fr}) - \mathbf{I}_{fr}\right]\hat{\mathbf{C}}_{fr}^\dagger\right)\overline{\mathbf{U}}^{-T}\\
 &= \left[\overline{\mathbf{S}}^{-1} + \mathbf{C}_{fr}\left(n_{FD}(\boldsymbol{\Lambda}_{fr}) - \mathbf{I}_{fr}\right)\mathbf{C}_{fr}^\dagger\right]\ .\label{eq:projDMPart1}
\end{align}
From the last equation, it appears that only the $N_{fr}$ largest Ritz-values are required to evaluate the Fermi-Dirac operator. The density matrix is the inverse of the overlap matrix plus a rank-$N_{fr}$ correction in which the largest eigenvectors of $\hat{\overline{\mathbf{H}}}$ appear. In a large system, $N_{fr}$ is generally much smaller than $N_{oc}$ and is more of less constant with respect to the system size. For example, our tests show that $N_{fr} \sim 8-32$ whereas $N_{oc} \sim 8,000-12,000$. Since only a few eigenvectors are required, iterative eigensolvers can be used to compute $\mathbf{C}_{fr}$.
Like the Rayleigh-Ritz algorithm, this partial Rayleigh-Ritz procedure works whether the subspace $\boldsymbol\Phi$ is orthonormal to begin with or not. It is generally faster to orthonormalize it inside the Rayleigh-Ritz procedure and not ahead of it. The partial Rayleigh-Ritz procedure is summarized in Algorithm \ref{alg:PRR}.

\begin{algorithm}
\caption{Partial Rayleigh-Ritz procedure}\label{alg:PRR}
\begin{algorithmic}
\Procedure{PartialRayleighRitz}{$\mathbf H$,$\boldsymbol\Phi$}
\State Compute $\overline{\mathbf{H}} = \boldsymbol\Phi^\dagger \mathbf H \boldsymbol\Phi$
\If{$\boldsymbol\Phi^\dagger \boldsymbol\Phi \neq \mathbf{I}$}
\State Compute $\overline{\mathbf{S}} = \boldsymbol\Phi^\dagger \boldsymbol\Phi$
\Else
\State $\overline{\mathbf{S}} = \mathbf{I}$
\EndIf
\State Diagonalize $\hat{\overline{\mathbf{H}}}\hat{\mathbf{C}}_{fr} = \hat{\mathbf{C}}_{fr}\boldsymbol\Lambda_{fr}$
\State Compute $\boldsymbol\Phi = \boldsymbol\Phi\overline{\mathbf{U}}^{-1}$
\State Compute $\overline{\mathbf{P}} = \mathbf{I} + \hat{\mathbf{C}}_{fr}\left(n_{FD}(\boldsymbol{\Lambda}_{fr}) - \mathbf{I}_{fr}\right)\hat{\mathbf{C}}_{fr}^\dagger$
\State \textbf{return} $\boldsymbol\Phi,\overline{\mathbf{P}}$
\EndProcedure
\end{algorithmic}
\end{algorithm}

To close, the partial Rayleigh-Ritz algorithm differs from the traditional Rayleigh-Ritz algorithm in two ways from a computational perspective. Firstly, in the Rayleigh-Ritz algorithm, the current Chebyshev filtered subspace is multiplied by the eigenvectors of the projected eigenvalue problems, a general matrix. In the partial Rayleigh-Ritz algorithm, the current Chebyshev filtered subspace is multiplied by the inverse of the Cholesky factor of the overlap matrix, a triangular matrix. In the case where the Chebyshev filtered subspace is already orthonormal, nothing needs to be done. This halves the computational cost associated with updating the subspace. In exact arithmetic, this could even be avoided altogether, but in practice the basis vectors of the subspace $\mathbf{\Phi}^k$ become linearly dependent and they must be periodically orthonormalized. Whether this is necessary can be monitored by looking at the condition number of the overlap matrix. Another distinction is that only a few eigenvalues and eigenvectors of the projected eigenvalue problem are necessary to build the one-particle density matrix and obtain the electronic density for the KS-DFT. This reduces the computational complexity associated with the diagonalization from $\mathcal{O}(N^3)$ to $\sim\mathcal{O}(N^2)$.

Furthermore, we argue that the partial Rayleigh-Ritz algorithm is easier to parallelize than the Rayleigh-Ritz algorithm. Parallelizing the eigenvalue problem Eq.(\ref{eq:projGV}) is tantamount to re-implementing ScaLAPACK routines or some other linear algebra library for distributed memory computers, a daunting task physicists wish to avoid. In contrast, in pRR it suffices to parallelize matrix products of the type $\hat{\overline{\mathbf{H}}}u$ to parallelize the eigenvalue problem in Algorithm \ref{alg:PRR}. The parallel matrix-vector routine can then be used in a partial diagonalization algorithm such as LOBPCG\cite{Knyazev01}.

\section{The RESCU Implementation}\label{sec:imp}

In this section, we report the implementation of the KS-DFT solver RESCU. The code is written in MATLAB and includes interfaces to MPI libraries, ScaLAPACK, CUDA, cuSPARSE and LibXC written in C and compiled into MEX-files.

The Kohn-Sham equation is discretized on a uniform Cartesian grid. High-order finite-differencing is used to discretize the differential operators. We generally use $\mathcal{O}(h^{16})$ stencils (49-point stencils) where $h$ is the grid resolution. Beyond that the Vandermonde system of equations determining the stencil coefficients becomes ill-conditioned. Moreover, the entries delimiting the stencil become negligibly small and there is no gain in trying to increase the accuracy further. Matrix representations of the first and second order differential operators are generated for each coordinate. Differencing matrix-vector products take the following form
\begin{align}
\label{eq:Ddiscrete}
{\partial^{(n)} \over \partial x_1^{(n)}}f(x_1^i,x_2^j,x_3^k) = \sum_l(\mathbf{D}_1^{(n)})^{i,l}f(x_1^l,x_2^j,x_3^k)
\end{align}
Here, the derivative is taken with respect to the first coordinate. More generally, $\mathbf{D}_j^{(n)}$ is a $n^{th}$ order discrete differential operator operating along the $j^{th}$ coordinate and $x_i^j$ is the $i^{th}$ grid point along the $j^{th}$ coordinate. Such products can be implemented as matrix-matrix products by making the array $f$ into a matrix in which the first dimension runs over the differentiated coordinate and the second dimension the other coordinates. $N$-dimensional finite-difference operators are Kronecker products of the form
\begin{align}
\label{eq:Krodiff}
{\partial^{(n)} \over \partial x_j^{(n)}} = \bigotimes_{i=1}^{j-1}\mathbf{I}_i\otimes\mathbf{D}_j^{(n)}\bigotimes_{i=j+1}^{N}\mathbf{I}_i
\end{align}
where $\mathbf{I}_i$ is the identity operator along the i$^{\text{th}}$ dimension. The matrix representation of the operator defined in Eq.(\ref{eq:Krodiff}) need not be built explicitly. It suffices to perform certain manipulations on the function array, array dimension permutation and transposition for instance, to make equation \ref{eq:Ddiscrete} into a matrix product. We implement gradients and Laplacians applications as these particular Kronecker products as we found this was most efficient. We have compared it against using the multi-dimensional (sparse) Laplacian, using the stencils directly and using Fourier transforms among others. For large systems it may be advantageous to put the differencing matrices $\mathbf{D}_j^{(n)}$ in a sparse format. A sparsity of roughly 0.05 was observed to be the turning point. For example, for an $\mathcal{O}(h^{16})$ stencil it would be advantageous to use sparse differential operators if the number of points along a dimension is larger than 300.

We use a pseudopotential set generated from the Troullier-Martins scheme\cite{TM91,NanoAcademic} and use the Kleinman-Bylander representation\cite{KB82} to model the atomic cores. A core correction is added as prescribed in Ref. \onlinecite{Louie1982} for elements in which the core shells overlap significantly with the valence shell. The set was developed for the NAO quantum transport package Nanodcal\cite{NanoAcademic} and it includes double-zeta polarized atomic orbitals. The Hartree potential of the spherically symmetric valence atomic orbital charge is added to the pseudopotentials, and hence screens long range Coulomb tails. Corrections to the Hartree potential are calculated by solving the Poisson equation for the deviation from the neutral atom density. Fourier transforms are used in periodic systems and sine transforms are used to diagonalize the finite-difference Laplacian in Dirichlet problems.

Our software implements the time-saving double-grid technique of Tomoya and Kikuji\cite{Tomoya99,Tomoya05}. It is used to compute certain integrals, such as the projection of the Kohn-Sham states onto the Kleinman-Bylander projectors, at a lower cost. The Kohn-Sham states are typically smoother than the pseudopotentials. The idea is to express the Kohn-Sham states on a coarse grid and interpolate them on a finer grid used for the pseudopotenials when needed, saving memory and time as a result. A few exchange-correlation functionals have been implemented in MATLAB: PW92\cite{PW92} (LDA), PBE\cite{PBE} (GGA) and MBJ\cite{MBJ} (mGGA). More exchange and correlation functionals are available from LibXC\cite{Marques12}. The interface from MATLAB to LibXC is written in C and allows us to use most LDA, GGA and meta-GGA functionals implemented in the library.

The Chebyshev filtering technique originally proposed by Zhou \textit{et al}. is significantly impeded by the initial subspace generation which requires solving for all Kohn-Sham eigenstates. In a recent paper, the authors show that starting from a random subspace and performing a few filtering steps is sufficient to obtain a suitable initial subspace\cite{Zhou14}. In our implementation, yet another option is available: a single- or double-zeta atomic orbital basis is used as an initial subspace. The lowest energy levels of that subspace are found by partly diagonalizing the projected Hamiltonian and then a quasi-minimal initial subspace is constructed. Moreover, it is possible to reach convergence to a prescribed accuracy in the NAO basis before transposing the calculation to real space. This provides a more robust and quick convergence and alleviate significant computational cost. Our tests show that even using a single-zeta atomic orbital basis can generally give a good initial subspace.

Certain elements with $d$ and $f$ electrons have hard pseudopotentials and the resolution of the grid must be increased accordingly. It is generally easier to perform calculations at a low resolution since the convergence rate of non-linear accelerators generally deteriorates with respect to system size as shown by Lin and coworkers in Ref.\onlinecite{Lin13-precond}. We have implemented a ``multi-grid" Kohn-Sham solver: it solves the equation on a coarse grid first and gradually refines the grid until some target resolution is reached. After completing the calculation at a given resolution, the code interpolates the pseudopotentials and neutral atom electronic density and interpolates the current approximation of the Kohn-Sham subspace on the refined grid. The density is then updated and a new self-consistent cycle is initiated. We implemented a few mixing schemes to accelerate the convergence of the density or the effective potential. In particular, we implemented Broyden mixing as proposed by Srivastava in Ref.\onlinecite{Srivastava84} and Johnson mixing\cite{Johnson88} as presented by Kresse and Furthm\"uller in Ref. \onlinecite{KresseMix}.

\begin{figure}
\centering

\null\hfill \subfigure[\ Tall process grid \label{fig:tallproc}]{
  \includegraphics[height=6cm]{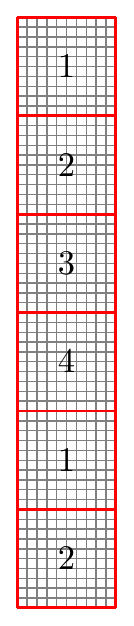}
}
\hfill
\subfigure[\ Flat process grid \label{fig:flatproc}]{
\includegraphics[height=6cm]{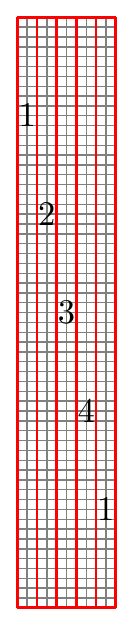}
}
\hfill
\subfigure[\ Square process grid \label{fig:sqrproc}]{
\includegraphics[height=6cm]{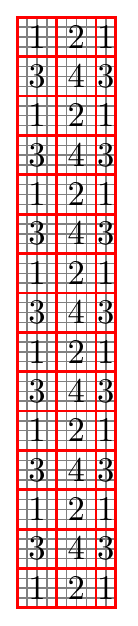}
}
\hfill\null

\caption{The 2D block cyclic distribution is a general scheme used to distribute arrays in RESCU. The array distribution for different block sizes and process grids is depicted. The numbers indicate the rank of the process holding the submatrices.}
\label{fig:procgrid}
\end{figure}

The parallelization is done with MPI and ScaLAPACK. MATLAB also naturally takes advantage of Intel's MKL threading capabilities. The MPI-only implementation is based on the MPI+ScaLAPACK implementation, and hence we refer to ScaLAPACK in the description that follows. Our implementation uses a 2D block cyclic distribution to scatter the arrays across many processes. This is a quite general distribution scheme as far as matrices are concerned. The distribution depends on four parameters: two specifying the size of the blocks, two specifying the size of the process grid. Some examples of a $60\times 10$ matrix shared between 4 processes are shown in Fig. \ref{fig:procgrid}. Fig. \ref{fig:tallproc} shows a matrix split in $10\times 10$ blocks distributed on a $4\times 1$ process grid, which we refer to as a tall process grid. In contrast, a flat process grid is distributed as shown in Fig. \ref{fig:flatproc}, where the blocks are $60\times 2$ submatrices. Finally, a matrix distributed on a square process grid is depicted Fig. \ref{fig:sqrproc}, where the blocks are $4\times 4$ submatrices. The block size must be chosen large enough to limit communication between processes but small enough to yield a good load balance. The performance is not that sensitive to block size in our experience; a block size between 16 and 256 is usually efficient on an InfiniBand network. It is much more sensitive to the shape of the process grid however. For certain operations, the computational time can vary as much as 100\%. It is more often than not favorable to take the time to redistribute optimally an array using PDGEMR2D before performing an operation. In the following description, we indicate which one of a tall, flat or square process grid is likely to yield the best performance.

The load associated with interpolating spatial variables such as potentials, densities and atomic orbitals is distributed according to a spatial partitioning which corresponds to a tall process grid (Fig. \ref{fig:tallproc}). A good load balance is achieved since the number of real space grid points is always by far superior to the number of processes and the communication cost is negligible since interpolation is a local operation.

A significant computational cost is associated with Hamiltonian-wavefunction products. Such products occur in the Lanczos solver which is used to compute the eigenspectrum upper bound (see Section \ref{sec:chebfilt}). The cost associated with the Lanczos solver is marginal but it can be parallelized by using ScaLAPACK with a few (e.g. four) processes or by threading the operations. Filtering the Kohn-Sham subspace is computationally expensive in comparison. We found that parallelizing over the Kohn-Sham states is often the most efficient approach since the processes do not need to communicate during the filtering procedure. We thus distribute the subspace $\mathbf{\Phi}_k$ array using a flat process grid (Fig.\ref{fig:flatproc}) prior to that operation. This is adequate as long as the number of electrons in the simulated system is larger than the number of processes. It is more difficult to achieve a good load balance in large systems with relatively few electrons such as `hollow' molecules and 2D crystals. In this case, a parallelization scheme as described in Ref.\onlinecite{Zhou06-2} will likely perform better. One other option is to use less processes but use threading to parallelize the Hamiltonian-wavefunction products.

Another computational bottleneck comes from orthonormalizing the Kohn-Sham subspace. Cholesky orthonormalization can be used, but it is performed by the more robust QR factorization routines PDGEQRF and PDORGQR if required. We have observed that matrix decomposition and diagonalization routines generally perform best on arrays split in square blocks and distributed on a square process grid (Fig.\ref{fig:sqrproc}). Consequently, we redistribute the subspace as it yields a good performance in our tests. A significant cost is associated with calculating the projected Hamiltonian and the overlap matrix. A tall process grid offers the best performance for this operation when using MPI or ScaLAPACK. The Rayleigh-Ritz procedure is completed by calling PDSYEV or PDSYGVX depending on whether an orthonormalization of the subspace has taken place before. In the Rayleigh-Ritz procedure, PDSYGVX finds all the eigenvectors. If the partial Rayleigh-Ritz procedure is used, then PDSYEVX or PDSYGVX is invoked to find the few required eigenvectors. We have also modified the ARPACK\cite{Lehoucq98} interface provided with MATLAB and Knyazev's MATLAB implementation of LOBPCG\cite{Knyazev01} to use the parallel matrix-vector products mentioned at the end of section \ref{sec:PRR}.

\section{Numerical Tests}\label{sec:num}

\begin{table*}[t]

\begin{tabular}{c c c c c c c c c}
\hline\hline
System &\# of atoms (e$^{-}$) & $N_x$ & $N_y$ & $N_z$ & Subspace (L) & Method & \# cores & Time (hrs) \\
\hline
Si & 5,832 (23,328) & 140 & 140 &140 & 11,672 & RS & 256 & 5.52 \\

Al & 4,000 (12,000) & 110 & 110 & 110 & 8,044  & RS & 64 & 5.09 \\

Al & 8,788 (26,364) & 141 & 141 & 141 & 17,596  & RS & 256 & 23.88 \\

Cu & 1,372 (15,092) & 156 & 156 & 156 & 8,058  & RS & 256 & 9.12 \\

DNA-H$_2$O & 5,399 (14,596) & 170 & 168 & 148 & 7,314 & RS & 256 & 9.62 \\

Si & 13,824 (55,296) & 247 & 247 & 247 & 55,296 & AO & 64 & 6.43 \\

Cu & 5,324 (58,564) & 267 & 267 & 267 & 95,832 & AO & 256 & 13.42 \\

\hline \hline
\end{tabular}
\caption{Some of the largest physical systems solved by KS-DFT with RESCU. The number of electrons in the system is indicated in the parentheses beside the number of atoms. The vector $[N_x,N_y,N_z]$ gives the numbers of points used along each dimension. For the AO method, $[N_x,N_y,N_z]$ is the size of the real space grid used to project the orbitals and calculate the density. It is also the grid by which the Poisson equation is solved for the Hartree potentials. The subspace dimension $L$ corresponds to the linear dimension of the eigenvalue problem. In the method column, RS stands for real space and AO for numerical atomic orbital.  The time is the total wall-clock time to converge the entire KS-DFT computation. More details about the computation are found in section \ref{sec:num}.}
\label{tb:largebench}
\end{table*}

Our numerical tests are performed at McGill University's Centre for High Performance Computing (HPC). We use nodes that consist of two Intel E5-2670 processors (8-core, 2.6 GHz, 20MB Cache) and 128 GB of DDR3 memory. The internode communication link is InfiniBand QDR. We use OpenMPI 1.8.3 and ScaLAPACK 2.0.2. In our tests, we set the number of processes equal to the number of cores and turn off threading. The timings reported below are wallclock times. The results for some of the largest systems are compiled in table \ref{tb:largebench}.

\subsection{Real Space RESCU}
\label{ssec:RSRESCU}

\subsubsection{Partial Rayleigh-Ritz}
\begin{figure}
\centering
\includegraphics{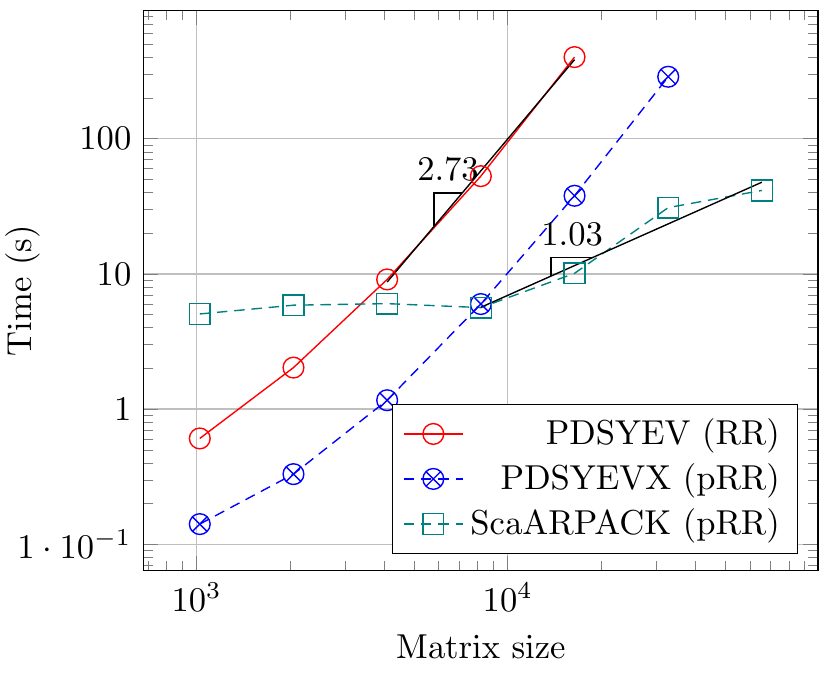}
\caption{Time as a function of matrix size for full and partial diagonalization (16 eigenpairs). The timings for the ScaLAPACK routines PDSYEV and PDSYEVX, used in the Rayleigh-Ritz and partial Rayleigh-Ritz algorithms, are represented by the circles and crossed circles respectively. The timings for the parallelized ARPACK function (ScaARPACK) are denoted by the squares.}
\label{fig:diagscaling}
\end{figure}

As mentioned earlier, the partial Rayleigh-Ritz algorithm is most competitive when the cost of the diagonalization taking place in the Rayleigh-Ritz procedure becomes significant. We evaluate the potential gain associated with diagonalization by benchmarking the ScaLAPACK routines PDSYEV, PDSYEVX and our parallel version of ARPACK, which we call ``ScaARPACK" here. We seek the 16 largest eigenvalues of random matrices of varying sizes. This is a typical number of buffer states required in the partial Rayleigh-Ritz algorithm in the simulation of gapped systems (e.g. Si). The time is averaged over 10 randomly generated symmetric matrices for each size. The ScaLAPACK routines use an $8\times 8$ process grid and $32\times 32$ blocks. ScaARPACK uses a homemade function that carries out the parallel matrix-vector products. The matrix is scattered according to a $64\times 1$ process grid and $16\times 16$ blocks. We have diagonalized matrices up to linear size 16,384, 32,768 or 65,536 with PDSYEV, PDSYEVX or ScaARPACK respectively. The results are plotted in Fig. \ref{fig:diagscaling}. The scaling is $\mathcal{O}(N^{2.7})$ for the ScaLAPACK routines and almost linear for ScaARPACK. This is better than theoretical asymptotic scaling in all cases. This reflects the importance of the communication cost in the case of ScaLAPACK and the large overhead of our implementation in the case of ScaARPACK. Both ScaLAPACK routines share roughly the same scaling but the routine PDSYEVX is about an order of magnitude faster than PDSYEV since it stops when the wanted eigenpairs have been found. ScaARPACK starts winning over PDSYEVX when the matrix to be diagonalized is larger than $8,000 \times 8,000$. The speed-up for the diagonalization becomes substantial when the number of electrons is equal to or greater than 32,000 (after accounting for spin degeneracy).

\subsubsection{Subspace Initialization}
Next, we demonstrate the importance of the quality of the initial subspace on convergence when using Chebyshev filtering in lieu of an eigensolver. In Ref. \onlinecite{Zhou06-2}, the authors suggest starting from a relatively accurate set of eigenvectors of the Hamiltonian. It was later demonstrated that using a few Chebyshev accelerated steps of the power method on a random initial subspace was sufficient to simulate many systems more efficiently\cite{Zhou14}. We thus compare the atomic orbital (NAO) initialization against the Chebyshev filtering initialization as described in Ref.\onlinecite{Zhou14}. Single-zeta atomic orbitals are interpolated on the real space grid and the Rayleigh-Ritz procedure is performed using the resulting subspace to obtain the initial subspace. In that sense, we are performing a one shot NAO calculation and then we project the result on the real space grid. For the Chebyshev filtering initialization, 4 Chebyshev filtering steps are used.

\begin{figure}
\centering
\includegraphics[width=\columnwidth]{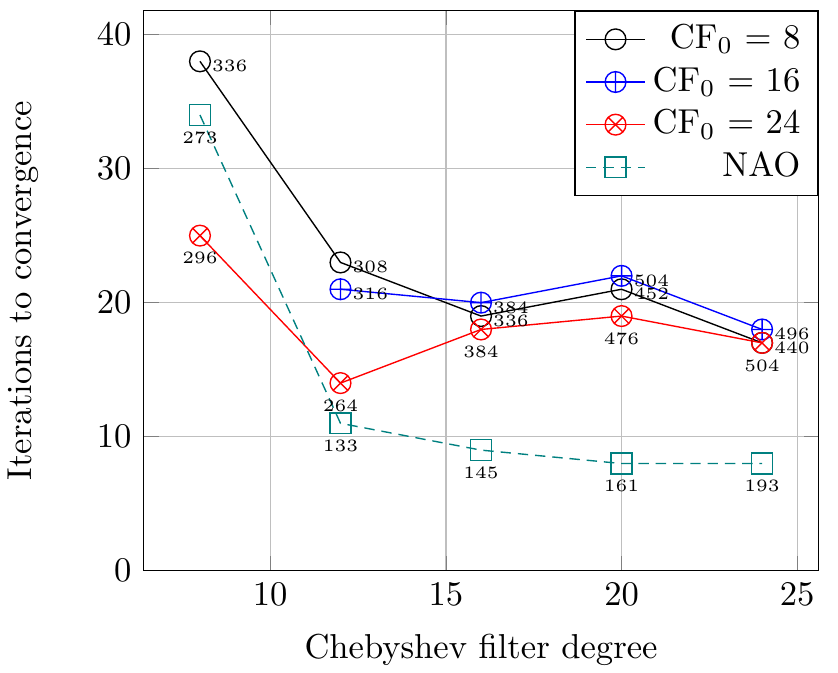}
\caption{Number of iterations to converge the density of a unit cell containing 216 Si atoms as a function of self-consistent Chebyshev filter degree for different initialization techniques. CF$_0$ stands for the degree of the Chebyshev filter used in the initialization of the subspace $\mathbf{\Phi}_k$ (4 filtering steps are used). NAO stands for initialization from the solution of a single-zeta atomic orbital basis. The total number of Hamiltonian-wave function products is written by the data points. The circle, ``plus'' circle and ``times'' circle marks are for CF$_0$ = 8, CF$_0$ = 16 and CF$_0$ = 24 respectively. The square marks are the the NAO initialization.}
\label{fig:cmpinit}
\end{figure}

In order to compare the methods, the density of a unit cell comprising 216 silicon atoms is calculated. The number of iterations to convergence as a function of the Chebyshev filter degree (CF$_\text{SCF}$) used in the self-consistent loop is plotted in Fig. \ref{fig:cmpinit}. We use 16 buffer states, Broyden acceleration with a mixing fraction of 0.3 and the convergence criteria are ${\|\rho_k - \rho_{k-1}\|\over N} < 10^{-5}$, ${\|E_k - E_{k-1}\|\over \|E_k + E_{k-1}\|} < 5\times 10^{-6}$. The number of iterations to converge the density for CF$_0 = 16$ and CF$_\text{SCF}$ = 8 is missing as the density did not converged within 100 iterations. CF$_\text{SCF} = 8$ also yields the worst performance for all initialization methods. This illustrates that the robustness is partly determined by the degree of the Chebyshev filter. NAO initialization leads to a faster convergence for all values of CF$_\text{SCF}$ except for CF$_\text{SCF} = 8$ in which case the CF$_0 = 24$ initialization leads to the smallest iteration count. However, the number of Hamiltonian-subspace products remains superior to the NAO case as indicated beside the marks in Fig. \ref{fig:cmpinit}. Otherwise, the degree of the filter in the initialization step does not seem to impact convergence much in the present test. For CF$_\text{SCF}$ larger than 12, the number of self-consistent steps to converge stagnates and the number of Hamiltonian-subspace products increases more of less linearly accordingly. Using NAO initialization, the inflation of the number of Hamiltonian-subspace products is not as important because the number of self-consistent steps keeps decreasing although not enough to compensate the cost of a high-degree filter. The optimal filter degree remains 12 for all methods for this system. We thus recommend using NAO initialization as it may converge faster and it is appreciably cheaper than Chebyshev filtering initialization in large systems by virtue of the localized character of the atomic orbitals (see Fig. \ref{fig:RESCUscalingSi} and Fig. \ref{fig:RESCUscalingSiLCAO}). The initial subspace can also be improved by using a multiple-zeta basis.

\subsubsection{Bulk Silicon Supercells}

\begin{figure}
\centering
\includegraphics[width=\columnwidth]{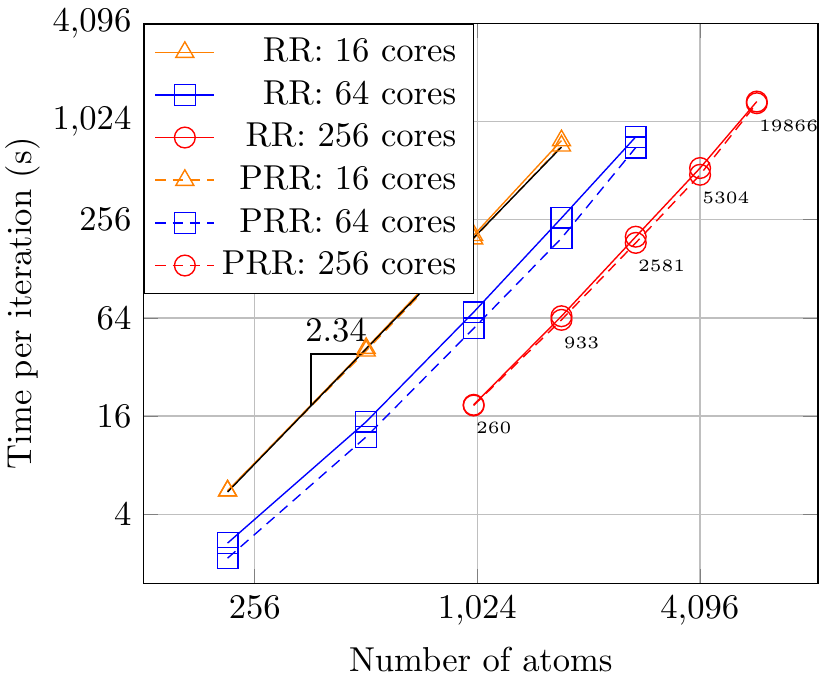}
\caption{Time per self-consistent step as a function of the number of Si atoms. The scaling with respect to the number of atoms is approximately $\mathcal{O}(N^{2.3})$. The triangles, squares and circles represent data points for calculations performed by 16, 64 and 256 cores respectively. The solid and dashed lines are for the Rayleigh-Ritz algorithm and the partial Rayleigh-Ritz algorithm. For the 256-core results, the numbers by the data points (red open circles) are the total wall-clock time for converging the entire KS-DFT run.}
\label{fig:RESCUscalingSi}
\end{figure}

Having tested various mathematical procedures, we now turn to physical systems of interest. Unit cells of silicon of varying size are simulated to test the performance of RESCU, the partial Rayleigh-Ritz algorithm and the Rayleigh-Ritz algorithm. The states are assumed to be spin-degenerate and 8 buffer states are included to separate the occupied and unoccupied spectra as explained in Section \ref{sec:chebfilt}. Kleinmann-Bylander projectors up to angular momentum $L=1$ are used. The grid spacing lower bound is set to 0.66 Bohr which corresponds to an energy cutoff of 300 eV. The differential operators are generated using 16$^{th}$ order stencils.
The exchange-correlation terms are computed using Perdew and Wang's version of LDA\cite{PW92} as implemented in LibXC. We use 15 steps of the Lanczos algorithm to calculate the eigenspectrum upper bound and a Chebyshev filter of degree 16.  Broyden mixing with a mixing fraction of 0.3 and a history of 20 is employed to accelerate convergence. The convergence criteria are again ${\|\rho_k - \rho_{k-1}\|\over N} < 10^{-5}$, ${\|E_k - E_{k-1}\|\over \|E_k + E_{k-1}\|} < 5\times 10^{-6}$.
For the partial Rayleigh-Ritz benchmark, we use the partial diagonalization capabilities of ScaLAPACK (PDSYGVX) and we find 12 hole eigenvalues (i.e. 8 buffer states plus 4 occupied states). Many of these parameters can be optimized further to yield a better performance: the parameters used here are by no means optimal - we simply used sensible values based on experience - but they already give impressive performance. Although this is suboptimal, we also keep the parameters constant while varying the number of processes to get a consistent and fair comparison.

In Fig. \ref{fig:RESCUscalingSi}, the time per self-consistent step is plotted as a function of the number of atoms. The calculations were carried out using 16, 64 and 256 cores. We could go up to 5,832 Si atoms before running out of memory (the next cubic Si supercell contains 8,000 Si atoms). The total time to convergence (in seconds) for the 256-core runs is written by the data points in Fig.\ref{fig:RESCUscalingSi}. We observe that, at to those sizes, the partial Rayleigh-Ritz algorithm leads to marginal gains. There are many reasons to this. Most importantly, the pRR gains are  masked by the large cost of projecting the Hamiltonian into the filtered subspace and computing the overlap matrix. Hence, even though pRR is always faster (see Fig.\ref{fig:diagscaling}), it gives marginal gains for this particular test. We shall discuss this issue in more details in section \ref{sec:discussion}. We stress that pRR is measured against the highly efficient parallel linear algebra library ScaLAPACK. In the case where no such library is available, pRR provides a convenient way to parallelize the computation of the projected density matrix and can lead to substantial time savings in smaller physical systems. Finally, we note that the computational time scales consistently as $\mathcal{O}(N^{2.3})$ for all processor counts. We also highlight that the parallelization efficiency approaches 100\% as the number of atoms in the system increases.

\subsubsection{Bulk Aluminium Supercells}

\begin{figure}
\centering
\includegraphics[width=\columnwidth]{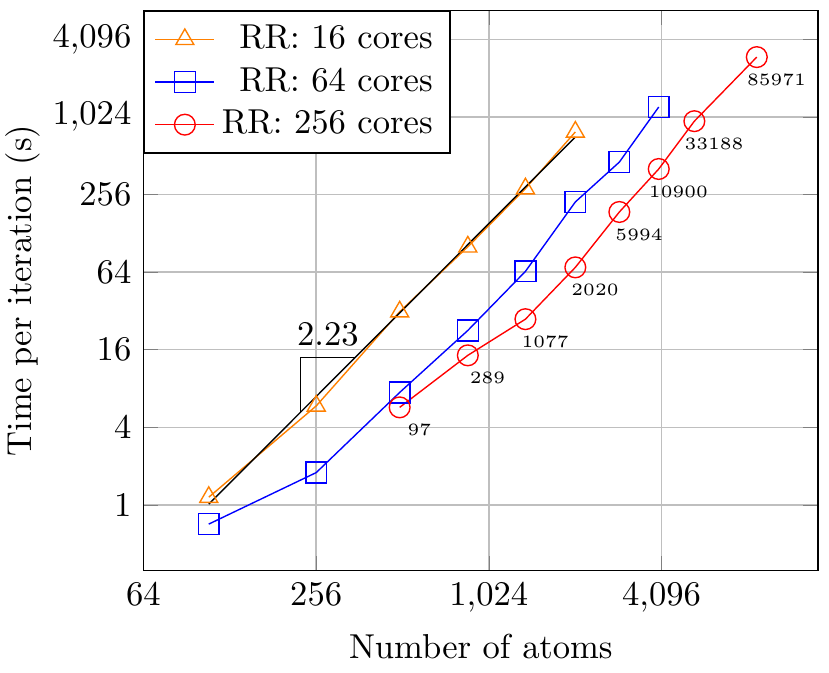}
\caption{Time per self-consistent step as a function of the number of Al atoms. The scaling with respect to the number of atoms is approximately $\mathcal{O}(N^{2})$. For the 256-core results, the numbers by the data points (red open circles) are the total wall-clock time for converging the entire KS-DFT run.}
\label{fig:RESCUscalingAl}
\end{figure}

We now turn to a metallic system for our second test: aluminium supercells. Again, we perform the calculations using 16, 64 and 256 cores. The number of valence electrons per atom is 3 and the system is assumed to be spin-degenerate. The spatial resolution is 0.7 a.u. and the differential operator accuracy is $\mathcal{O}(h^{16})$. We compute the states occupancies using the Fermi-Dirac distribution with a temperature of 1,000K. The computation is performed within the LDA using the routines XC\_LDA\_X and XC\_LDA\_C\_PW from LibXC as in the previous benchmark. We chose a Chebyshev filter of degree 16. The number of buffer states varies with respect to system size. This is necessary to open a sizeable gap between the occupied and unoccupied states as explained in section \ref{sec:chebfilt}. In the present test, we set the number of extra states to 10\%-30\% the number of ions. Finally, we used Johnson-Kerker mixing with a mixing fraction of 0.5 and a minimal mixing fraction of 0.05. The time per self-consistent step as a function of the number of Al atoms is displayed in Fig. \ref{fig:RESCUscalingAl}. The computational cost per self-consistent step scales almost quadratically ($\mathcal{O}(N^{2.2})$) with respect to the number of atoms. With 64 cores, the largest supercell simulated contained 4,000 Al atoms and the density and total energy where converged to one part in 10$^{5}$ in slightly over 18,000 seconds using 256 cores. A supercell containing 8,788 Al atoms is handled with 256 cores. After 29 iterations and almost 24 hours of computation, the residuals are the following: ${\|\rho_k - \rho_{k-1}\|\over N} \simeq 10^{-4}$, ${\|E_k - E_{k-1}\|\over \|E_k + E_{k-1}\|} \simeq 5\times 10^{-8}$. The density convergence criterion is not as restrictive as one used in the other benchmarks. We note, however, that it is already constrained enough for band structure calculation purposes as shown in Fig.\ref{fig:albs} below.

\subsubsection{Bulk Copper Supercells}

We briefly report input parameters and results for another metal test: copper supercells. In this test we used 256 cores. The number of valence electrons per atom is 11 and the system is assumed to be spin-degenerate. The spatial resolution is 0.3 a.u. and the differential operator accuracy is $\mathcal{O}(h^{16})$. We compute the states occupancies using the Fermi-Dirac distribution with a temperature of 100K. The computation is performed within the LDA using the routines XC\_LDA\_X and XC\_LDA\_C\_PW from LibXC as in the previous benchmark. We chose a Chebyshev filter of degree 16. In the present test, we set the number of extra states to 50\% the number of ions. Finally, we used Johnson-Kerker mixing with a mixing fraction of 0.25 and a minimal mixing fraction of 0.1. The largest supercell simulated contained 1,372 Cu atoms and the density and total energy where converged to one part in 10$^{5}$ in 32,843 seconds using 256 cores as reported in table \ref{tb:largebench}.

\subsubsection{DNA molecule in water}

\begin{figure}
  \centering
  \includegraphics[width=\columnwidth]{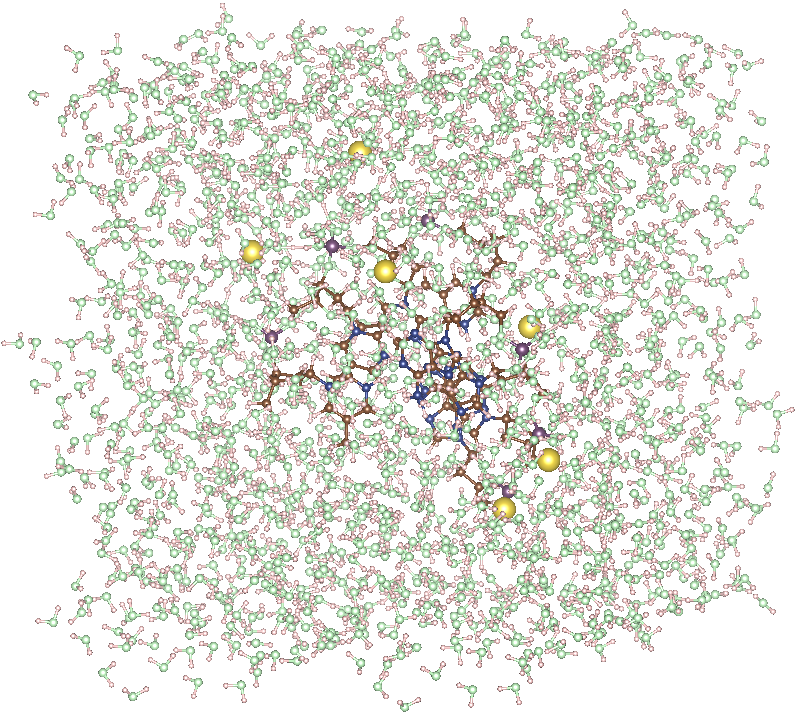}
  \caption{DNA molecule solvated in 1,713 water molecules.}
  \label{fig:dna}
\end{figure}

As another example of testing on physical systems, RESCU is applied to calculate the electronic structure of a solvated DNA structure (5'-AAAA-3') which is a completely disordered system. The initial structure is obtained from minimizing the structure with the molecular modeling package AMBER 11. The DNA structure is initially charge neutralized with counterions by 6 Na$^+$ and solvated with 1,713 TIP3P water molecules\cite{Jorgensen}. The system is depicted in Fig. \ref{fig:dna}. The tetragonal simulation domain has dimensions $44.5\times44.4\times39.1 {\AA}^3$ and periodic boundary conditions. We use a resolution of 0.25 Bohr. A total of 5,399 atoms (14,596 electrons) were simulated using 256 cores in the setup described above. We use the LibXC XC\_GGA\_X\_OPTPBE\_VDW exchange functional and XC\_GGA\_C\_OP\_B88 correlation functional. The Laplacian is discretized using $\mathcal{O}(h^{16})$ stencils. We used a 16$^{th}$ order Chebyshev filter and 16 buffer states. The convergence criterion is ${\|\rho_k - \rho_{k-1}\|\over N} < 10^{-5}$ and we mix the density using the Pulay method with a mixing fraction of 0.1. We initialize the subspace using a single-zeta atomic orbital basis set with an angular momentum cutoff $L=1$. The electronic density converged in 20 steps which took a total of 34,638 seconds for an average time of 1,732 seconds per self-consistent step. Finally, we find that the gap between the highest occupied molecular orbital (HOMO) and the lowest unoccupied molecular orbital (LUMO) shrinks to $0.6$eV - the gap for the isolated DNA structure without water is $2.0$eV - which indicates that the solvent plays an important role in the optical properties of wet DNA\cite{Singh}. We shall present the comparison in a forthcoming article\cite{lei}.

\begin{figure}
\centering
\includegraphics[width=\columnwidth]{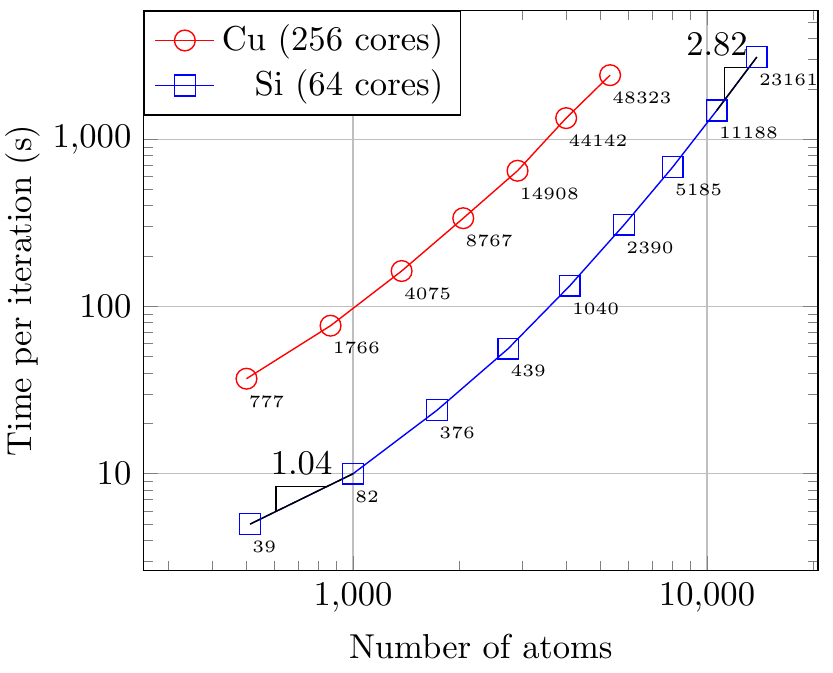}
\caption{Time per self-consistent step as a function of the number atoms. The circles are for the Cu benchmark which is performed using a double-zeta atomic orbital basis with angular momentum cutoff $L=2$ and 256 cores, each having 8GB of DDR3 memory. The squares are for the Si benchmark which is performed using a single-zeta atomic orbital basis with angular momentum cutoff $L=1$ and 64 cores, each having 4GB of DDR3 memory. The scaling with respect to the number of atoms is linear $\mathcal{O}(N^{1.04})$ for smaller systems and gradually ramps up to cubic $\mathcal{O}(N^{2.82})$. The numbers by each data point is the total wall-clock time for converging the entire KS-DFT run.}
\label{fig:RESCUscalingSiLCAO}
\end{figure}

\subsection{Atomic Orbital RESCU}
\label{ssec:AORESCU}

To demonstrate the capabilities of the atomic orbital method implementation in RESCU, we perform the same benchmark as above (bulk Si supercells) using numerical atomic orbitals. We use the same processors (E5-2670) equipped with half the memory this time (4GB/core). We perform the benchmark with 64 cores only.  We also use a real space resolution of 0.5 a.u. A single-zeta basis with angular momentum cutoff $L=1$ was used such that there are 4 atomic orbitals per atom. The results are plotted as blue squares in Fig. \ref{fig:RESCUscalingSiLCAO}. The largest simulated Si supercell comprises 13,824 Si atoms and the KS-DFT was converged in 23,161 seconds. Even at that size, we did not run into any memory issue since the atomic orbital basis yields sparse matrices. For relatively modest supercells, RESCU scales linearly. The scaling deteriorates as the supercell size grows since the eigenvalue problem accounts for a larger and larger proportion of the computational cost. The scaling gradually becomes cubic since we are treating the projected eigenvalue problem as a dense eigenvalue problem. Many methods evoked in Section \ref{sec:algorithms} could improve the efficiency of NAO calculations further. Since diagonalization performance is so important in this test, we mention that a square processor grid and 64$\times$64 blocks are used in the block cyclic distribution of the Hamiltonian and overlap matrices.

We have also performed NAO computations for bulk copper supercells. For this benchmark, we use 256 cores with 8GB of memory per core. We use a real space resolution of 0.25 a.u and a double-zeta basis with angular momentum cutoff $L=2$ (18 atomic orbitals per atom). We have simulated supercells including up to 5,324 Cu atoms (58,564 electrons). The density and total energy were converged in roughly 12 hours (33 iterations). The time per self-consistent step as a function of the number of Cu atoms is indicated by the red circles in Fig. \ref{fig:RESCUscalingSiLCAO}. The number of atomic orbitals per atom is 4.5 times larger than in the Si benchmark and the number of cores 4 times larger. Consequently, the time per step for the Cu system is roughly an order of magnitude larger than the time per step for the Si system. The time per iteration scales similarly, i.e. it goes from linear in the 1,000 atom range to cubic in the 10,000 atoms range. The total time scales worse here as larger (metallic) systems tend to take more iterations to converge.

\subsection{Accuracy}

\begin{figure}
\centering

\subfigure[\ Si band structure \label{fig:sibs}]{
  \includegraphics[width=\columnwidth]{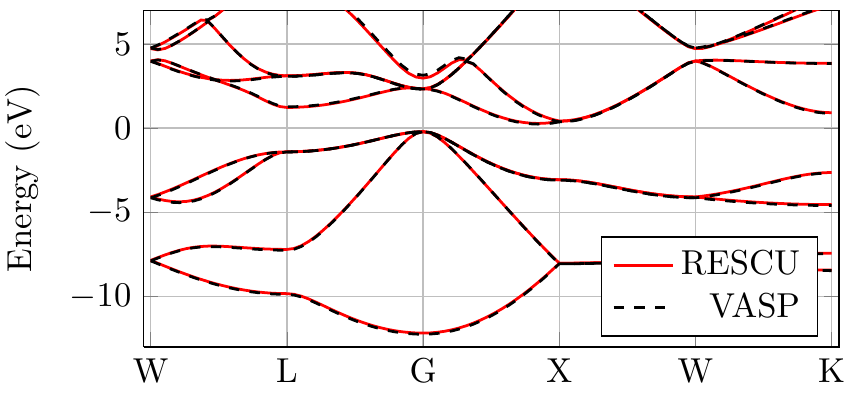}
}

\subfigure[\ Cu band structure \label{fig:cubs}]{
\includegraphics[width=\columnwidth]{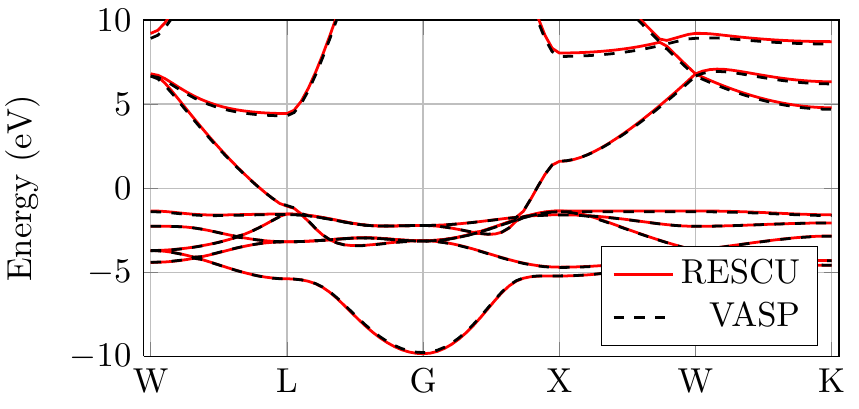}
}

\subfigure[\ Al band structure \label{fig:albs}]{
\includegraphics[width=\columnwidth]{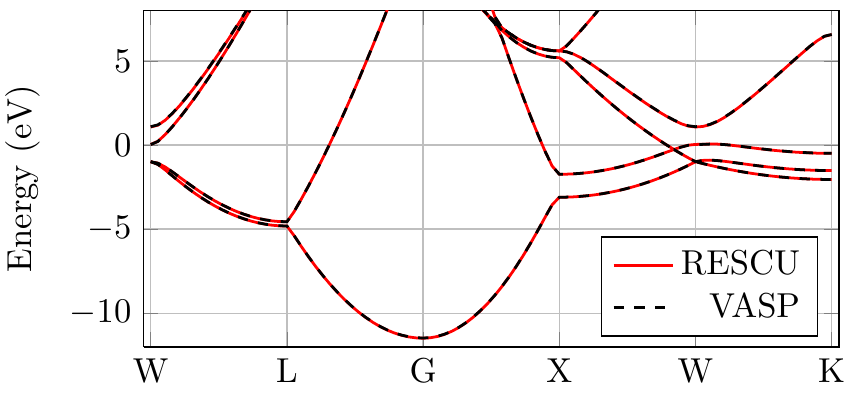}
}

\caption{Comparisons of band structures obtained by VASP and RESCU. The band structures agree to a high accuracy indicating that the resolution used in our benchmark was sufficient for the purpose of band structure calculations.}
\label{fig:bs}
\end{figure}

\begin{figure}
\centering

\subfigure[\ GaAs band structure \label{fig:sibs}]{
  \includegraphics[width=\columnwidth]{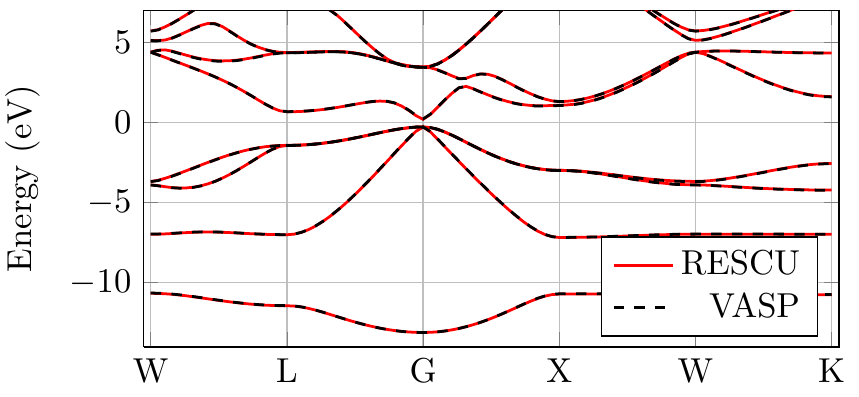}
}

\subfigure[\ MgO band structure \label{fig:cubs}]{
\includegraphics[width=\columnwidth]{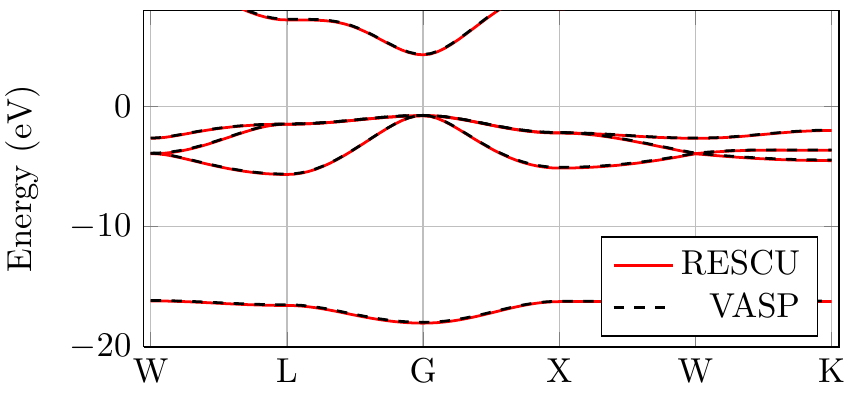}
}

\caption{Comparisons of the band structures of the GaAs and MgO compounds obtained by VASP and RESCU.}
\label{fig:bs2}
\end{figure}

To verify the accuracy of the electronic structures calculated in our benchmark, we calculated the band structures of Al, Si and Cu using the VASP package (albeit using a primitive cell) and compared it with the band structures calculated from the density of the largest supercell simulated. The overlayed RESCU and VASP band structures are displayed  in Fig.\ref{fig:bs}. The agreement is impressive considering that these methods are different in many respects. In particular, the PAW method is used in VASP and pseudopotentials are used in RESCU. We have also calculated band structures of compounds by both RESCU and VASP, results for the semiconductor GaAs and the insulator MgO are presented in Fig. \ref{fig:bs2}. Again, the agreement between the two methods is excellent. This demonstrates the precision of the RESCU method for band structure calculations.

A systematic approach to comparing DFT codes has been introduced by Lejaeghere et al. \cite{Cottenier2014} in 2014. Its essence is to calculate the $\Delta$-functional (see Eq.\ref{eq:Delta} below) for the total energy versus volume (E vs V) equation of states (EOS) of elemental crystals. The equilibrium volume $V_0$, the bulk modulus $B_0$ and the derivative of the bulk modulus $B_1$ can be extracted from the EOS using a third-order Birch-Murnaghan fit\cite{Birch1947}. The value of $\Delta$ is thus a sensible measure of the agreement of the structural and mechanical predictions of two DFT solvers. The $\Delta$-functional is defined as follows
\begin{equation}
\label{eq:Delta}
\Delta(E^a,E^b) = {1\over |\Omega|}\int\limits_\Omega dV \left(E^a(V) - E^b(V)\right)
\end{equation}
where $\Omega$ is an interval, $E^a(V)$ is the E vs V EOS for code $a$ and $E^b(V)$ is the E vs V EOS for code $b$. The interval $\Omega$ is chosen as $[0.94(V_0^a+V_0^b)/2, 1.06(V_0^a+V_0^b)/2]$.

We have calculated the value of $\Delta(E^{RESCU},E^{WIEN2k})$ for a number of  elements, where $E^{RESCU}$ is the EOS calculated by RESCU and $E^{WIEN2k}$ is the EOS calculated by WIEN2k \cite{Schwarz2003}. The WIEN2k EOS are provided in the supplemental content of Ref. \onlinecite{Cottenier2014}. In RESCU, we used a grid spacing of 0.14 Bohr and 6750$/N$ k-points for $N$-atoms unit cells. The charge convergence criterion is $10^{-5} e$ per valence electron and the energy convergence criterion is $10^{-5}$ Hartree per valence electron. A Fermi-Dirac smearing with a temperature of 800 K is used. The computation is performed within the GGA using the routines XC\_GGA\_X\_PBE and XC\_GGA\_C\_PBE from LibXC. Again, in the RESCU calculation the atomic cores are modeled by Troullier-Martins pseudopotentials \cite{TM91,NanoAcademic}. The obtained $\Delta$ values are listed in Table \ref{tb:delta} and the differences between the EOS are quite reasonable in all cases. These $\Delta$ values are comparable to those obtained with the electronic structure package AbInit (using the Troullier-Martins pseudopotential) and WIEN2k\cite{deltacodesdft}. Many codes include all elements from H to Rn except those between lanthanum and ytterbium in their test, but we leave such an exhaustive test to a future opportunity.

We end this subsection by emphasizing that, while using a more efficient solution process to solve the KS-DFT equation, there was no accuracy-degrading approximation in RESCU and the accuracy tests presented here strongly demonstrate its quality.
\begin{table}
\begin{tabular}{ll|ll}
\hline\hline
Element & $\Delta$ (meV/atom) & Element & $\Delta$ (meV/atom)\\
\hline
H  & 0.864 & Ca & 2.013\\
Be & 7.080 & Cu & 7.758\\
Mg & 1.392 & Rh & 18.627\\
Al & 0.434 & Pd & 15.581\\
Si & 3.608 & Ag & 11.286\\
P  & 7.291 & Cs & 1.082\\
S  & 6.830 & &\\
\hline\hline
\end{tabular}
\caption{The $\Delta(E^{RESCU},E^{WIEN2k})$ values for thirteen elements.}
\label{tb:delta}
\end{table}

\section{Further Discussions}
\label{sec:discussion}

\begin{figure}
\centering
\includegraphics[width=\columnwidth]{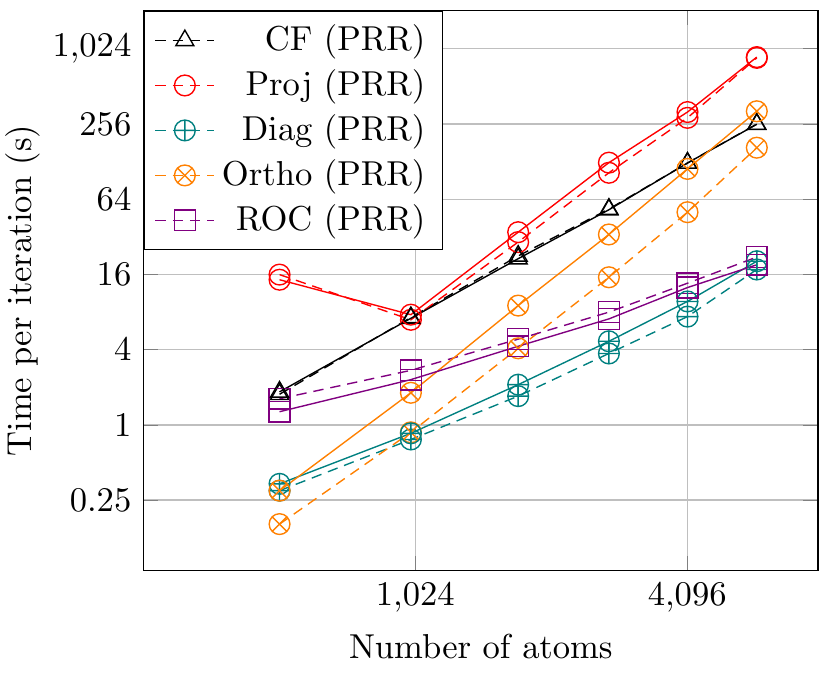}
\caption{Time per self-consistent step as a function of the number of Si atoms using 256 processors. The full line link data points for the Rayleigh-Ritz algorithm and the dashed lines link data points for the partial version.}
\label{fig:RESCUdecompSi}
\end{figure}

Having demonstrated the power of RESCU by solving the KS equation for thousands of atoms - both insulating and metallic, both ordered and disordered, and on a modest computer cluster - we now discuss the principal bottlenecks of the real space method in RESCU. To this end, we use the results of the Si benchmark in Section \ref{sec:num} for the discussion. We have plotted the time taken by the computationally intensive operations of Table \ref{tb:KSscaling} as a function of the number of atoms in Fig. \ref{fig:RESCUdecompSi}: they are the Chebyshev filtering, the Hamiltonian projection onto the filtered subspace, the diagonalization of the projected pencil, the orthonormalization or/and computation of the Ritz vectors, and we add to this list the residual timing (ROC) which includes the remaining time. The timings for one self-consistent step performed by 256 cores are reported in Fig. \ref{fig:RESCUdecompSi}. Both RR and pRR algorithms yield similar timings in all parts of the computation except for the orthonormalization. In solving the generalized eigenvalue problem, the eigensolver computes the Cholesky factor of the overlap matrix to reduce the generalized eigenvalue problem to a standard form. In the partial Rayleigh-Ritz algorithm, this factor is reused to orthonormalize the filtered subspace. In the standard Rayleigh-Ritz algorithm, there is no point in doing so since the eigenvectors for the generalized eigenvalue problem will directly provide the Ritz vectors which are orthonormal by construction. But the latter matrix is a general one whereas the former is triangular, and hence the factor of two speed up observed in Fig. \ref{fig:RESCUdecompSi}. The diagonalization is faster in the partial Rayleigh-Ritz procedure, but it is not too significant as it is as large as the efficiency fluctuations observed in our computation tests. There are a few explanations to this. Firstly, according to the benchmark in Fig.\ref{fig:diagscaling}, direct diagonalization is relatively cheap for matrices smaller than $10,000\times 10,000$ which is about the size of the matrices in our largest system. Secondly, in a relatively simple system of bulk silicon, the projected matrix pencil $\bar{\mathbf{H}} - \lambda\bar{\mathbf{S}}$ is more easily diagonalizable than the random matrices used in the benchmark described above since it is closer to being diagonal. Moreover, it becomes closer and closer to being ``diagonal'' as the electronic density - and the Kohn-Sham invariant subspace - approaches its fixed point.  The scaling of the residual timing appears linear and that of the Chebyshev filtering is quadratic. The main bottlenecks are the projection and the orthonormalization which scale almost cubically. This originates from the complexity of matrix-matrix multiplication which is $\mathcal{O}(N^3)$ or $\mathcal{O}(N^{2.8})$ depending on matrix size and implementation.

The main bottleneck comes about because of the massive amount of data required to encode the subspace $\mathbf{\Phi}_k$. In order to improve the method described in this work, it is crucial to achieve some sort of subspace compression. One way to address this issue is to calculate a localized basis for the filtered subspace as done by Motamarri \textit{et al}. in Ref. \onlinecite{Motamarri14}, where the authors show that localizing the basis is key to achieving a subquadratic computational scaling, in particular when evaluating the Hartree-Fock exchange functional. Tangentially, subspace compression is crucially needed because the memory requirement scales as $\mathcal{O}(N^2)$. Indeed, the RESCU method is so far more limited by the memory requirement than the computational requirement as evidenced in the Si benchmark. Unfortunately, it appears that, for the reported tests, the subspace ``resists'' localization as it approaches convergence and performing part of the computation with a dense subspace is still required. In a similar spirit, RESCU can use multiple-zeta numerical atomic orbital bases to solve the Kohn-Sham equations. If the memory requirements are not too severe, the NAO solutions are projected into real space and the energy may be further minimized using Chebyshev filtering. In summary, future research should focus on the following points: a vector space spanning the Kohn-Sham occupied subspace must be computed efficiently, storing such a vector space should require $\mathcal{O}(N)$ memory (the data must be compressible somehow) and the projected density matrix should be efficiently computable. In any event, even with the existing bottlenecks, RESCU has already achieved impressive computational efficiency in solving the KS-DFT problem.

\section{Conclusions}\label{sec:conclusion}

In this work, we have presented a powerful Kohn-Sham DFT solver, RESCU. The goal of the RESCU method is to predict electronic structure properties of systems comprising many thousands of atoms using moderate computer resources. The computational efficiency is gained by exploiting four routes. First, in real space, Chebyshev filtering is used to expedite the computation of an invariant Kohn-Sham subspace in large systems. This approach essentially exploits the fact that when the Hamiltonian is not yet converged, one does not need to solve the KS equation extremely accurately. Second, we developed a NAO-based method to efficiently generate a good initial subspace which is necessary in the Chebyshev filtering paradigm.  Third, by judiciously analyzing various parts of the KS-DFT solution procedure, RESCU gains efficiency by delaying the $O(N^3)$ scaling to large $N$; and our tests showed that RESCU scales as $O(N^{2.3})$ up to the several thousand atoms level. Fourth, RESCU gains efficiency by various numerical mathematics and, in particular, we introduced the partial Rayleigh-Ritz algorithm and showed it leads to efficiency gains for systems comprising more than 10,000 electrons. The RESCU code is implemented in MATLAB such that it provides a convenient prototyping and development environment. It is also easily installed on many platforms and architectures. Finally, we mention in passing that we have also implemented total energy and force calculation methods into RESCU, but we reserve the discussion of structural relaxation using RESCU for the future.

We demonstrated that the RESCU method could perform large scale KS-DFT computations using computer resources ranging from 16 to 256 cores. At the 5,000-15,000 atoms level, there are many important material physics problems to be investigated and we wish to report them in the near future. From the method development point of view, to deal with even larger systems, we find it is essential to compress the Kohn-Sham subspace to achieve better computational effectiveness and to relieve computer memory requirements. Sparse matrix representations can alleviate the problem to some extent but do not solve it entirely. We think the solution may lie in the hierarchical matrix approximations which are data-sparse structures for dense matrices. We wish to present these efforts in the future. Finally, it is tempting and interesting to extrapolate the RESCU ability to much larger systems - using current supercomputers. However, we caution that a naive extrapolation may not work since for much larger $N$, the computational burden shifts toward the parts of the algorithm that scale as $O(N^3)$. We believe these are important topics of future research.

\textbf{Acknowledgement.}
We gratefully acknowledge financial support by NSERC of Canada and FQRNT of Quebec (H.G.). We thank Dr. Eric Zhu and Dr. Lei Liu for their help on pesudopotentials and LCAO basis sets; Dr. Langhui Wan and Dr. Kevin Zhu for helping us on computational issues related to ScaLAPACK. We thank McGill HPC, Calcul Qu\'ebec and Compute Canada for computation facilities which made this work possible.

\bibliographystyle{prsty}
\bibliography{PartialRR2}

\end{document}